\UseRawInputEncoding
\documentclass[letter,prapplied, amsmath,amssymb,
 reprint,
 author-year,%
superscriptaddress,
longbibliography
]{revtex4-2}

\usepackage{natbib}
\usepackage[export]{adjustbox}

\usepackage{float}
\usepackage{graphicx}
\usepackage{amsmath,amsthm,amssymb,amsfonts}
\usepackage{dcolumn}
\usepackage{bm}
\usepackage{gensymb}
\usepackage[normalem]{ulem}
\usepackage{hyperref}
\usepackage[all]{hypcap} 
\usepackage[caption=false]{subfig}
\graphicspath{ {images/} }
\usepackage[colorinlistoftodos]{todonotes}
\usepackage{wasysym}
\usepackage{array}
\usepackage{xcolor}
\usepackage[utf8]{inputenc} 
\usepackage{float}
\usepackage{siunitx}
\usepackage{braket}

\newcolumntype{C}{>{\centering\arraybackslash}X} 

\usepackage{cleveref}

\AtBeginDocument{%
    \newwrite\bibnotes
    \def\bibnotesext{Notes.bib}
    \immediate\openout\bibnotes=\jobname\bibnotesext
    \immediate\write\bibnotes{@CONTROL{REVTEX41Control}}
    \immediate\write\bibnotes{@CONTROL{%
    apsrev41Control,author="08",editor="1",pages="1",title="0",year="1"}}
     \if@filesw
     \immediate\write\@auxout{\string\citation{apsrev41Control}}%
    \fi
}%

\crefname{equation}{Eq.}{Eqs.}
\Crefname{equation}{Equation}{Equations}
\crefname{figure}{Fig.}{Figs.}
\Crefname{figure}{Figure}{Figures}
\crefname{section}{Sect.}{Sects.}
\Crefname{section}{Section}{Sections}
\crefname{table}{Table}{Tables}
\crefname{appsec}{Appendix}{Appendices}
\graphicspath{{Figures/}}

\usepackage{color, colortbl}
\definecolor{Gray}{gray}{1}
\definecolor{orange}{rgb}{1,0.97,0.9}
\definecolor{cyan}{rgb}{0.92,1,1}
\definecolor{red}{rgb}{1,0.9,0.9}
\definecolor{purple}{rgb}{0.5, 0.0, 0.5}

\usepackage{soul}

\usepackage{xr}
\usepackage{hyperref}

\usepackage[caption=false]{subfig}

\begin{document}

\title{Fast Sideband Control of a Weakly Coupled Multimode Bosonic Memory}

\author{Jordan Huang}
\email{jah499@scarletmail.rutgers.edu}
\affiliation{Department of Physics and Astronomy, Rutgers University, Piscataway, NJ 08854, USA}
\author{Thomas J. DiNapoli}
\affiliation{Department of Physics and Astronomy, Rutgers University, Piscataway, NJ 08854, USA}
\author{Gavin Rockwood}
\affiliation{Department of Physics and Astronomy, Rutgers University, Piscataway, NJ 08854, USA}
\author{Ming Yuan}
\affiliation{Pritzker School of Molecular Engineering, University of Chicago, Chicago, Illinois 60637, USA}
\author{Prathyankara Narasimhan}
\affiliation{Department of Physics and Astronomy, Rutgers University, Piscataway, NJ 08854, USA}
\author{Eesh Gupta}
\affiliation{Department of Applied Physics, Stanford University, Stanford, California 94305, USA}
\author{Mustafa Bal}
\affiliation{Superconducting Quantum Materials and Systems Division, Fermi National Accelerator Laboratory (FNAL), Batavia, IL 60510, USA}
\author{Francesco Crisa}
\affiliation{Superconducting Quantum Materials and Systems Division, Fermi National Accelerator Laboratory (FNAL), Batavia, IL 60510, USA}
\author{Sabrina Garattoni}
\affiliation{Superconducting Quantum Materials and Systems Division, Fermi National Accelerator Laboratory (FNAL), Batavia, IL 60510, USA}
\author{Yao Lu}
\affiliation{Superconducting Quantum Materials and Systems Division, Fermi National Accelerator Laboratory (FNAL), Batavia, IL 60510, USA}
\author{Liang Jiang}
\affiliation{Pritzker School of Molecular Engineering, University of Chicago, Chicago, Illinois 60637, USA}
\author{Srivatsan Chakram}
\email{schakram@physics.rutgers.edu}
\affiliation{Department of Physics and Astronomy, Rutgers University, Piscataway, NJ 08854, USA}
\date{\today}

\begin{abstract} 
Circuit quantum electrodynamics (cQED) with superconducting cavities coupled to nonlinear circuits like transmons offers a promising platform for hardware-efficient quantum information processing. A critical challenge is developing control protocols that mitigate cavity errors arising from ancilla-induced decoherence, nonlinearities, and crosstalk. We address this by weakening the dispersive coupling while also demonstrating fast, high-fidelity multimode control by dynamically amplifying gate speeds through transmon-mediated sideband interactions. This approach enables transmon-cavity SWAP gates, for which we achieve speeds up to 30 times larger than the bare dispersive coupling.  Combined with transmon rotations, this allows for efficient, universal state preparation in a single cavity mode, though achieving unitary gates and extending control to multiple modes remains a challenge. In this work, we overcome this by introducing two sideband control strategies: (1) a shelving technique that prevents unwanted transitions by temporarily storing populations in sideband-transparent transmon states and (2) a method that exploits the dispersive shift to synchronize sideband transition rates across chosen photon-number pairs to implement transmon-cavity SWAP gates that are selective on photon number. We leverage these protocols to prepare Fock and binomial code states across any of ten modes of a multimode cavity with millisecond cavity coherence times. We demonstrate the encoding of a qubit from a transmon into arbitrary vacuum and Fock state superpositions, as well as entangled NOON states of cavity mode pairs\textemdash a scheme extendable to arbitrary multimode Fock encodings. Furthermore, we implement a new binomial encoding gate that converts arbitrary transmon superpositions into binomial code states in $\qty{4}{\micro\second}$ (less than $1/\chi$), achieving an average post-selected final state fidelity of $\qty{96.3}{\percent}$ across different fiducial input states. By using precalibrated transmon and sideband pulses, our work demonstrates multimode control with significantly reduced calibration requirements, enabling efficient unitary operations using sideband interactions in multimode cQED systems.

\end{abstract}

\pacs{Valid PACS appear here}
\maketitle

\section{Introduction}
Superconducting cavities coupled to nonlinear, ancillary circuits form a promising platform for quantum computing based on circuit quantum electrodynamics (cQED)~\cite{wallraff2004strong}. This architecture leverages the superconducting circuit's nonlinearity to manipulate quantum information stored in high-coherence cavity modes~\cite{reagor2016quantum}. When coupled to a multimode cavity, this architecture also allows multiplexed control of many cavity modes with only a few control lines~\cite{naik2017random,chakram2020seamless}. Universal control of a harmonic oscillator has been realized through a variety of control schemes~\cite{heeres2017implementing, eickbusch2021fast}. These control schemes have been extended to multiple oscillators to demonstrate entanglement~\cite{wang2016schrodinger, chakram2020multimode} and gate operations between multiple cavity modes, using both fixed-frequency transmons~\cite{rosenblum2018cnot,gao2019entanglement} and tunable coupler circuits~\cite{chapman2023high, lu2023high}. The harmonic oscillator's expanded Hilbert space and straightforward decoherence mechanisms---primarily photon loss---have also enabled demonstrations of hardware-efficient quantum error correction (QEC) using bosonic codes~\cite{ofek2016extending,hu2019quantum,campagne2020quantum, ni2023beating, sivak2023real}. 

The dispersive coupling to a nonlinear circuit, while crucial for implementing universal cavity control, introduces ancilla-mediated errors that are frequently the dominant error channels. In particular, the cavity modes inherit decay from the lossy, nonlinear circuit through the inverse-Purcell effect, setting a fundamental upper bound on cavity coherence---an effect that becomes increasingly significant as cavity lifetimes are extended~\cite{romanenko2020three, milul2023superconducting,oriani2024niobium, pietikainen2024strategies}. In multimode architectures, the always-on dispersive interaction and the nonlinearities inherited by the cavity modes introduce coherent crosstalk between modes~\cite{you2024crosstalk}, further contributing to gate infidelity and placing constraints on the scalability of these systems.

One approach to mitigating these challenges is to weaken the dispersive interaction, thereby reducing the magnitude of ancilla-mediated errors. However, this reduction in interaction strength also slows gate speeds in traditional control schemes that rely on photon-number-selective ancilla rotations~\cite{heeres2015cavity, heeres2017implementing, bretheau2015quantum, chakram2020multimode}. To address this trade-off, control strategies utilizing conditional cavity displacements have been developed. In these schemes, interaction rates are enhanced beyond the bare dispersive coupling by using large cavity displacements as an interaction switch~\cite{eickbusch2021fast, diringer2024conditional}. Another method for increasing gate speeds beyond the dispersive coupling strength is to use ancilla displacements to activate charge-driven sideband interactions~\cite{milul2023superconducting}. The transmon's quartic nonlinearity can activate various parametric processes~\cite{wallraff2007sideband}, such as cavity beamsplitters~\cite{pfaff2017controlled, gao2018programmable} and transmon-mode SWAP gates using two drive photons~\cite{mollenhauer2024high}. The fastest transmon-mediated sideband interaction for a given drive strength uses a single drive photon to convert two transmon excitations into a single photon in a coupled cavity mode~\cite{pechal2014microwave, rosenblum2018cnot}.

We use such charge-driven sidebands to realize a tunable Jaynes\textendash Cummings interaction (within the transmon's $\ket{g}$-$\ket{f}$ manifold) between the transmon and any target mode of a multimode cavity. Here, the coupling is modulated by the strength of the transmon drive. When combined with transmon rotations, this enables efficient, universal state preparation using the Law--Eberly protocol~\cite{law1996arbitrary}\textemdash originally developed to control the motional states of trapped ions~\cite{ben2003experimental} and later demonstrated in superconducting circuits~\cite{hofheinz2009synthesizing}. This protocol reverses the sequence that brings the target state to the vacuum one photon at a time and has a sequence length sub-linear in the maximum photon number of the target state. However, incommensurate sideband interaction rates across different photon-number states from the $\sqrt{n}$ bosonic enhancement pose significant challenges in implementing arbitrary gate operations~\cite{liu2021constructing}. Furthermore, calibrating the arbitrary transmon and sideband rotation angles required for a given unitary can be challenging, especially in the presence of large Stark shifts. This difficulty is analogous to that encountered when realizing target unitaries via optimal control pulses generated through algorithms such as GRAPE~\cite{khaneja2005optimal, heeres2017implementing}. In traditional dispersive control~\cite{heeres2015cavity} or with conditional displacement gates~\cite{eickbusch2021fast}, employing discrete families of pulses that can be independently calibrated and optimized offers advantages in both ease of implementation and robustness against classical control errors and quantum state-dependent shifts~\cite{fosel2020efficient, kudra2022robust}.

\begin{figure}
  \begin{center}
    \includegraphics[width= 0.5\textwidth]{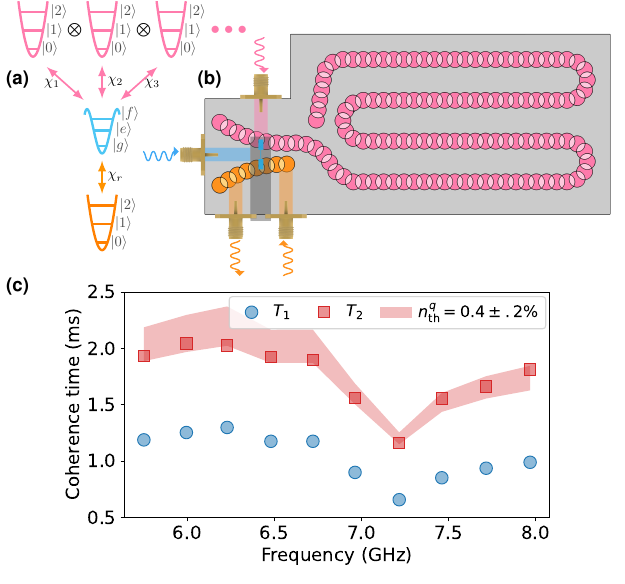}
    \caption{\textbf{Multimode bosonic memory device.} \textbf{(a)} Device architecture and \textbf{(b)} schematic depicting the multimode cavity-based quantum processor~\cite{chakram2020seamless, chakram2020multimode} comprised of a multimode cavity (pink) with cavity modes coupled to and controlled by a single transmon circuit (blue), which is measured using a readout cavity (orange). The cavities are fabricated using the flute method~\cite{chakram2020seamless, chakram2020multimode}, by drilling offset, overlapping, evanescent holes in a monolithic block of high-purity (5N5) aluminum. The hole locations are indicated in the schematic. \textbf{(c)} The measured coherence times of the storage cavity modes. The shaded regions indicate the inferred qubit thermal population that would give rise to the measured cavity $T_2$.}
    \label{fig1}
  \end{center}
\end{figure} 

In this work, we overcome these challenges by introducing two sideband control strategies: the first is a shelving technique that temporarily stores populations in the transmon $\ket{e}$ state so that they remain unaffected by sideband drives. The second method implements photon-number-selective sideband operations by matching sideband rates across chosen photon numbers, using the sideband drive detuning as a control knob. Using these strategies, we introduce a method of state preparation and unitary synthesis that is implemented through a discrete set of precalibrated transmon and sideband pulses. Using these techniques, we prepare binomial code states~\cite{michael2016new} across ten cavity modes in approximately $\qty{1}{\micro\second}$. We also implement unitary operations that encode a qubit from the transmon into vacuum and Fock state superpositions, as well as into entangled NOON states between chosen cavity mode pairs. Finally, we implement a new encoding unitary that transfers a transmon superposition into a binomial logical qubit in any target mode at a rate exceeding the dispersive shift. We demonstrate these multimode operations on a cavity memory with ten active modes with a weak dispersive coupling to a transmon ancilla. In this regime, we achieve SWAP operations that are $30$ times faster than the bare dispersive coupling.

\section{Multimode bosonic memory}

\begin{figure*}
  \begin{center}

    \includegraphics[width=\textwidth]{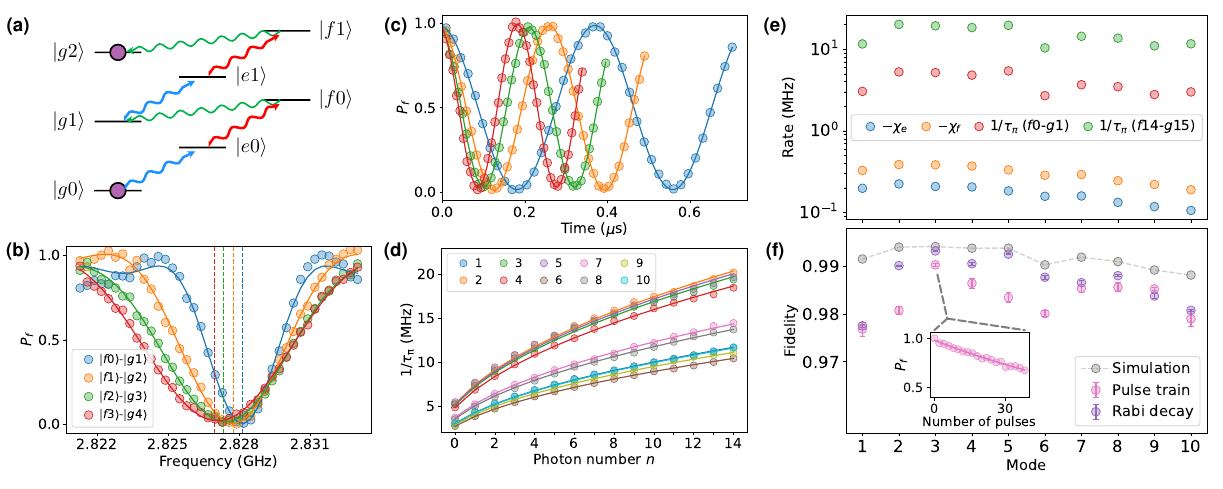}
    \caption{\textbf{Charge-driven sideband interactions in the multimode memory.} \textbf{(a)} Energy level diagram for a transmon and a single cavity mode illustrating how to climb the levels with transmon rotations (blue and red) and $\ket{f,n}$-$\ket{g,n+1}$ sideband transitions (green). \textbf{(b)} Spectroscopy and \textbf{(c)} and stimulated vacuum Rabi oscillations of $\ket{f,n}$-$\ket{g,n+1}$ transitions for mode 3 of the storage cavity. \textbf{(d)} Sideband Rabi oscillation rate as a function of photon number for the first ten modes of the multimode memory, up to a photon number of $n=14$. \textbf{(e)} Comparison of the maximum achieved SWAP rate for the $\ket{f0}$-$\ket{g1}$ and $\ket{f13}$-$\ket{g14}$ transitions against the bare dispersive interaction strengths for the $\ket{e}$ and $\ket{f}$ states. \textbf{(f)} The pink markers show the $\ket{f0}$-$\ket{g1}$ SWAP fidelities inferred from a pulse train experiment. The inset shows the pulse train experiment performed for mode three. The data is fit to an exponential decay as a function of pulse number to extract the fidelity. The purple markers show the $\ket{f0}$-$\ket{g1}$ SWAP fidelities inferred from fitting the $\ket{f0}$-$\ket{g1}$ oscillations to an effective Rabi model as in \cite{lu2023high}. The grey markers show the sideband interaction fidelities extracted from Lindblad master equation simulations.}
      \label{fig2}
  \end{center}
\end{figure*} 

Our system architecture and device schematic are shown in Fig.~\ref{fig1}(a) and (b). The device features a multimode rectangular waveguide cavity supporting several high-coherence modes, each with lifetimes ranging from $0.6$\textendash$\qty{1.3}{\milli\second}$, as shown in Fig.~\ref{fig1}(c). The cavity is fabricated from a monolithic block of high-purity (5N5) aluminum using the flute method, where the cavity volume is formed by drilling offset, overlapping, evanescent holes~\cite{chakram2020seamless}. The mode spectrum is set by the cavity dimensions, with the lowest mode around \qty{5.75}{\giga\hertz}. We realize a nearly evenly spaced mode spectrum, with separations of approximately \qty{250}{\mega\hertz}, by tapering the cavity height. The transmon is inserted at the taller end of the cavity with transmon-mode couplings chosen to realize dispersive shifts of $\chi \approx 100$\textendash$\qty{200}{\kilo\hertz}$. With the transmon's $T_1\approx\qty{56}{\micro\second}$, this leads to an inverse-Purcell limit of about \qty{60}{\milli\second} for the coherence times of the cavity modes due to their coupling to the lossier transmon. The transmon frequency is $f_q = \qty{4.606}{\giga\hertz}$, its dephasing time is $T_2 \approx\qty{66}{\micro\second}$, and the thermal population is $0.3 \pm \qty{0.1}{\percent}$. The transmon is also coupled to a separate flute-style cavity whose fundamental mode at $f_r = \qty{8.067}{\giga\hertz}$ is used for transmon readout. In addition to readout input and output drives, the system has direct transmon and multimode storage cavity drive lines.

\section{Fast charge-driven sideband interactions}

The quartic nonlinearity of the transmon mediates a sideband interaction between the states $\ket{f, n}$ and $\ket{g, n+1}$, where $n$ is the number of photons in the target cavity mode \cite{PhysRevA.91.043846}. This interaction is initiated by applying a charge drive on the transmon at the frequency difference between its $\ket{f}$ state and the cavity mode. While these states do not directly couple, an interaction is facilitated through virtual states which are coupled to $\ket{f, n}$ and $\ket{g, n+1}$ through the dressing from the transmon drive and the Jaynes--Cummings coupling with the target cavity mode (the virtual states being $\{\ket{e, n}, \ket{e, n+1}\}$ for small drive amplitudes). The resulting sideband interaction can be described as $g_{\text{sb},i}\left(\ket{g}\bra{f}b^\dag_{i} + \ket{f}\bra{g}b_{i}\right)$, where $b_{i}$ is the lowering operator for photons in cavity mode $i$. Its interaction rate, computed to lowest order in the nonlinearity, without making a rotating wave approximation for the transmon displacement, is given by
\begin{equation}
    g_{\text{sb},i} = \frac{\sqrt{2}(\omega_q + K + \Delta_i)}{2\omega_q + K + \Delta_i}\frac{\epsilon}{g_i}\chi_i \approx \frac{\epsilon}{\sqrt{2} g_i}\chi_i.
\end{equation}
Here, $\omega_q$ is the transmon's $\ket{g}$-$\ket{e}$ frequency, $K$ is its anharmonicity, $\Delta_i = \omega_q - \omega_{c,i}$ is the transmon-mode detuning, $g_i$ is the bare transmon-mode coupling, and $\chi_i$ is the dispersive shift of the transmon $\ket{e}$ state from addition of a photon in mode $i$. In this form, it is clear that the sideband interaction can be enhanced beyond $\chi$ with a strong drive $\epsilon$. Higher-order corrections to the sideband rate are discussed in Section S3 B of the Supplementary Information.

In conjunction with transmon rotations, the sideband interaction allows us to climb the ladder of Fock levels of any cavity mode, as schematically illustrated in the level diagram shown in Figure~\ref{fig2}(a). We measure the resonance frequency of the $\ket{f, n}$-$\ket{g, n+1}$ transition by performing sideband spectroscopy measurements after preparing the transmon in the $\ket{f}$ state. We start with the cavity in the ground state and locate the $\ket{f, 0}$-$\ket{g, 1}$ resonance. Next, we sweep the pulse duration to identify the $\pi$-pulse time corresponding to adding a single photon in the target cavity mode. This implements a transmon-cavity SWAP gate. The process is iteratively repeated for higher photon numbers, identifying the $\ket{f, n}$-$\ket{g, n+1}$  transitions and extracting the corresponding $\pi$ pulse times (shown in Figure \ref{fig2}(b) and (c)). We observe that the resonance frequency of the $n$-photon sideband is offset by $n\chi_f$ due to the dispersive shift of the $\ket{f}$ state. Additionally, the interaction rates exhibit a $\sqrt{n}$ bosonic enhancement with photon number as shown in Figure \ref{fig2}(d). Figure \ref{fig2}(e) compares the rate of the $\ket{f, 0}$-$\ket{g, 1}$ SWAP gate to the dispersive interaction ($\chi_e, \chi_f$) of the modes. We achieve SWAP gate times nearly 30 times faster than $\chi_e$ by using strong transmon drive strengths ranging from $0.1$\textendash$\qty{1.3}{\giga\hertz}$. At these drive strengths, the transition frequency has a significant drive-amplitude-dependent Stark shift (approximately $10$\textendash$\qty{50}{\mega\hertz}$). Here, the Floquet modes adiabatically connected to the undriven transmon states have significant support with higher transmon levels. To maximize the interaction fidelity, we use a smooth ramp (bump function) that is slow enough to allow adiabatic evolution of the undriven states to their corresponding Floquet modes and fast enough to induce high-fidelity population exchange. This corresponds to the ramp initializing the initial state into an equal superposition of Floquet modes at the final drive amplitude through a Landau-Zener-like transition (see Section~S7 of the Supplementary Information).

\begin{figure*}
  \begin{center}
    \includegraphics[width= 1.0\textwidth]{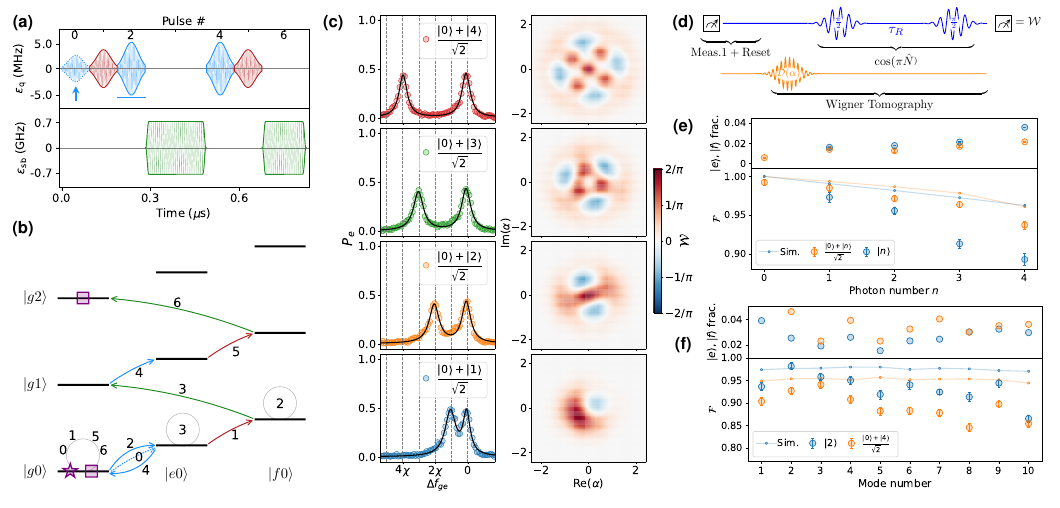}
    \caption{\textbf{Single-mode control using transmon rotations and sideband transitions.} \textbf{(a)} Pulse sequence for preparing $(\ket{0}+\ket{2})/\sqrt{2}$ in a target mode in $\sim 800$ ns, corresponding to $\approx \frac{1}{6\chi}$ for this mode (3). The arrow indicates the initial pulse used to prepare a transmon superposition, while the blue underline denotes a shelving pulse. An arbitrary superposition of $\cos\theta\ket{0}+e^{i\varphi}\sin\theta\ket{n}$ can be encoded by adjusting the initial transmon pulse rotation. \textbf{(b)} Energy level diagram representing the combined Hilbert space of the transmon and one of the cavity modes, illustrating the state transitions induced by the pulse sequence in (a). The purple star indicates the initial state, and the purple squares indicate the final superposition state. The numbers indicate the states occupied after each pulse in the sequence. The circles denote the state remaining unchanged by the corresponding pulse. \textbf{(c)} (Left) Photon number–resolved spectroscopy and (right) Wigner tomography of the cavity state after preparing $(\ket{0}+\ket{n})/\sqrt{2}$ for $n=1$–$4$, post-selected on the transmon being in $\ket{g}$. For the preparation of $\ket{n}$ in any cavity mode, the shelving pulses are omitted and the transmon is initialized in $\ket{e}$ with a $\pi_{ge}$ pulse. \textbf{(d)} The pulse sequence used for Wigner tomography with post-selection, comprising an initial transmon measurement followed by a standard Wigner tomography sequence that includes a cavity displacement and subsequent parity measurement. \textbf{(e)} (Top) Fraction of shots not in $\ket{g}$ in the first measurement as a function of $n$ following preparation of $\ket{n}$ and $(\ket{0}+\ket{n})/\sqrt{2}$. (Bottom) Corresponding fidelities obtained from Wigner tomography conditioned on the transmon being in $\ket{g}$. \textbf{(f)} (Top) Fraction of population not in $\ket{g}$ and (bottom) Wigner tomography fidelities post-selected on the transmon being in $\ket{g}$ after preparing binomial code states in any of the cavity modes.}
      \label{fig3}
  \end{center}
\end{figure*}

We characterize the fidelity of the sideband SWAP gate using an error amplification experiment where we measure the $\ket{f}$ state population following a train of $\ket{f0}$–$\ket{g1}$ $\pi$-pulses. An optimally calibrated sideband pulse results in a nearly exponential decay of the population, as shown in the inset of Fig.~\ref{fig2}(f), whereas calibration errors lead to non-exponential behavior due to coherent over- or under-rotations. We fit the exponential decay to extract a pulse fidelity, which includes errors from decoherence and leakage to higher transmon levels. To isolate the contribution of decoherence alone, we fit the decay of a long-time sideband Rabi oscillation to an effective Rabi decay model~\cite{lu2023high}. The fidelities extracted from both pulse train and Rabi decay measurements are shown in Fig. \ref{fig2}(f), along with a comparison with master equation simulations. The pulse-train fidelities range from approximately $97.5$\textendash$\qty{99}{\percent}$ and are found to be marginally lower than the Rabi decay fidelity, attributed to residual leakage errors caused by imperfections in the pulse ramps. The fidelities from Rabi decay are found to be consistent with master equation simulations that account for transmon and cavity decay and dephasing. In addition to decoherence and leakage errors, the pulses used in the gate protocols are susceptible to coherent errors arising from calibration drifts during the measurements. These drifts are attributed to drive-amplitude fluctuations, which lead to frequency errors due to the large Stark shifts. We mitigate this by performing iterative calibrations of the sideband frequencies and $\pi$-pulse times for all target photon numbers before each experiment. We detail the procedure followed to calibrate and characterize the sideband pulses in Section S4 of the Supplementary Information. At sufficiently high sideband drive strengths, which vary between modes, fidelities degrade due to transmon ionization~\cite{shillito2022dynamics}, imposing a fundamental limit on gate speed. To avoid this, we constrain the drive strength below this threshold.

We also measure the thermal population of the cavity modes by using a new sideband-based protocol that extends the standard protocol used for transmon thermal population measurements~\cite{jin2015thermal}. This is estimated by comparing the contrast of $\ket{f0}$-$\ket{g1}$ sideband rotations with and without initially preparing the transmon in $\ket{f}$, yielding cavity-mode thermal populations in the range of $0.3$\textendash$\qty{2}{\percent}$. To further reduce thermal excitations, we implement transmon reset using the $\ket{f0}$-$\ket{g1}$ sideband with our readout resonator, which has a short lifetime of approximately $\qty{600}{\nano\second}$. Combined with sidebands to the storage modes, we implement a sequential sideband reset protocol applicable to any storage mode. In this protocol, detailed in Section S6 of the Supplementary Information, the cavity state is reset one photon at a time by shuttling photons to the transmon $\ket{f}$ state and removing it through the readout mode. This increases our experimental repetition rate by two orders of magnitude ($\qty{7}{\milli\second}$ to $\qty{50}{\micro\second}$) and cools the cavity modes to a thermal population of $0.2$\textendash$\qty{0.7}{\percent}$, consistent with the expected thermal population of our readout resonator, as inferred from transmon dephasing.

\section{Single-mode sideband control}

We demonstrate single-mode control over any cavity mode by preparing both Fock states $|n\rangle$ and vacuum–Fock superpositions $\frac{|0\rangle+|n\rangle}{\sqrt{2}}$ by climbing the Jaynes–Cummings ladder with transmon rotations and sideband pulses. Fock states are prepared using a sequence of broadband $\pi_{ge}$ and $\pi_{ef}$ pulses followed by $\ket{f,n}$-$\ket{g,n+1}$ sideband $\pi$ pulses. To achieve superposition states, we employ our \textit{shelving} method—temporarily storing part of the state in the $|e\rangle$ manifold, which is transparent to the $\ket{f,n}$-$\ket{g,n+1}$ sidebands. We track the evolution of each component throughout the sequence, arranging it so that only one component is selectively transferred via sideband transitions while the others remain unaffected. Without shelving, the components would evolve at different rates due to the incommensurate sideband interaction rates across different photon numbers. Figure~\ref{fig3}(a) shows the pulse sequence used to prepare the superposition state $\frac{\ket{0}+\ket2}{\sqrt{2}}$, with shelving pulses indicated by blue underlines. Figure~\ref{fig3}(b) depicts the corresponding state evolution, with the numbers corresponding to the pulse number and the arrows indicating the superposition state at the end of each pulse. By adjusting the initial transmon pulse (indicated by the blue arrow), we note that the sequence implements a unitary operation that encodes transmon superpositions into arbitrary vacuum–Fock state superpositions, $\cos\theta|0\rangle+e^{i\varphi}\sin\theta\,|n\rangle$.

We characterize the prepared states by performing photon-number-resolved transmon spectroscopy to obtain the Fock-state populations of the cavity state. We also use Wigner tomography to fully characterize the cavity state density matrix. These protocols assume that the transmon is in the $|g\rangle$ state and initially disentangled from the cavity. However, decoherence and coherent calibration errors can lead to residual entanglement between the transmon and cavity modes at the end of gate operations. We mitigate the resulting uncontrolled state-characterization errors by measuring the transmon before state characterization and post-selecting on outcomes where the transmon is in $\ket{g}$. To minimize the effect of cavity decoherence during the initial post-selection measurement, we implement a fast integrated readout and reset pulse~\cite{mcclure2016rapid}, detailed in Section S8 C of the Supplementary Information. Figure~\ref{fig3}(c) (left) shows the photon-number-resolved spectroscopy of the transmon, while Figure~\ref{fig3}(c) (right) shows the corresponding reconstructed Wigner functions of the prepared superposition states up to $n=4$, both after post-selection. The pulse sequence used for Wigner tomography with post-selection is shown in Figure~\ref{fig3}(d). 

The fidelities for Fock ($|n\rangle$) and vacuum-Fock superposition states ($\frac{|0\rangle+|n\rangle}{\sqrt{2}}$) as a function of $n$ for mode three are shown in Figure~\ref{fig3}(e) (bottom), with the fraction of shots not in $\ket{g}$ shown on top. Similarly, the fidelities for preparing binomial code states in any of the ten modes of the cavity, post-selected on the transmon in $\ket{g}$, are shown in Figure~\ref{fig3}(f). The experimental results are compared with master simulations incorporating transmon and cavity decoherence. The discrepancy is attributed to a combination of imperfect calibration of the sideband pulses, and underestimating the fidelity of the state due to noise in the Wigner tomography measurements. The error bars for the fidelities are extracted by considering both the noise in the data and the uncertainty in the parity measurement calibration. The noise in the data leads to both a spread in the reconstructed fidelity as well as a systematic lowering. The calibration and error analysis for Wigner tomography are detailed in Section S8 D of the Supplementary Information. We note that the weak dispersive coupling results in a long duration for the parity measurement ($1/(2\chi)\sim \qty{2.5}{\micro\second}$), and the cavity-state measurement is affected by ancilla decoherence.  This can potentially be avoided in the weak coupling limit by instead measuring the characteristic function of the cavity state through conditional displacements~\cite{campagne2020quantum,eickbusch2021fast}. In our case, to obtain accurate state fidelities, we employ an error mitigation technique by rescaling the parity measurement to account for transmon decoherence during the parity measurement (see Section S8 A in the Supplementary Information). We also account for the reduction in contrast due to the finite bandwidth of the parity measurement.  

\begin{figure}
  \begin{center}
    \includegraphics[width= 0.5\textwidth]{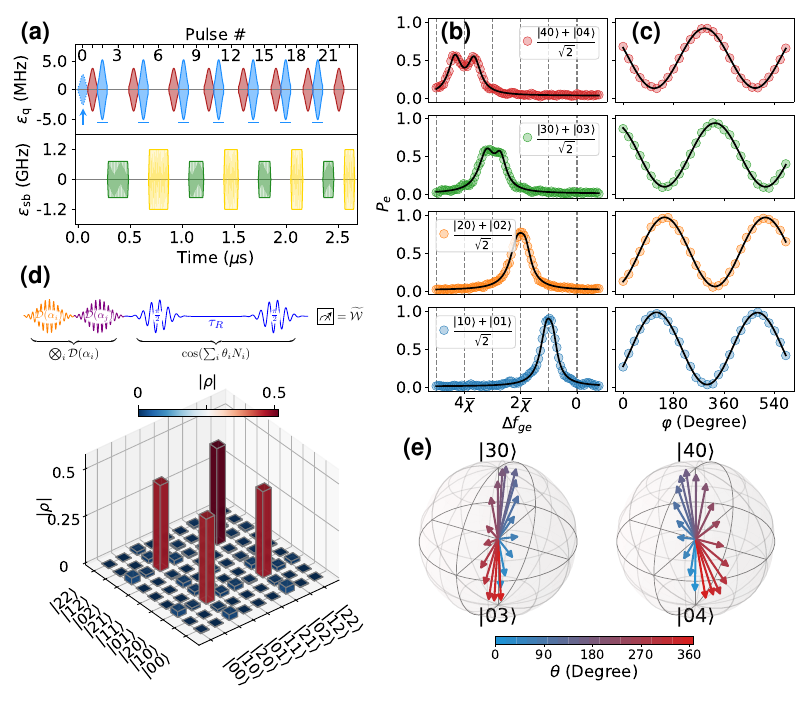}
    \caption{\textbf{Preparation and characterization of two-mode entangled states.} \textbf{(a)} Pulse sequence for preparing a NOON state with $N = 4$. The initial pulse preparing an arbitrary transmon $\ket{g}$-$\ket{e}$ superposition is indicated by an arrow while the blue underlines indicate the shelving pulses. \textbf{(b)} Photon-number-resolved spectroscopy of NOON states (between modes three and five) up to $N=4$. $\bar{\chi}$ is the average dispersive shift of the two modes. \textbf{(c)} Coherence of the NOON state measured using a $\pi/2$ pulse with varying phase after decoding the NOON state back to the transmon $\ket{g}$-$\ket{e}$ subspace. \textbf{(d)} The density matrix of a NOON state ($N=2$) extracted using multimode Wigner tomography \textbf{(e)} Tomography of $\ket{\psi} = \cos\theta\ket{N0} +e^{i\varphi}\sin\theta\ket{0N}$ versus $\theta$ within the NOON state subspace, for $N=3, 4$. This was measured by decoding the NOON state back into the transmon $\ket{g}$-$\ket{e}$ subspace and performing qubit tomography.}
    \label{fig4}
  \end{center}
\end{figure} 

\section{Multimode Entanglement}

We extend the protocol used to encode vacuum-Fock superpositions to multiple modes and prepare entangled NOON states of the form $\cos\theta\ket{N0} + e^{i\varphi}\sin\theta\ket{0N}$. The pulse sequence to implement this is shown in Figure~\ref{fig4}(a). Shelving pulses to the $\ket{e}$ manifold (underlined in blue) ensures that sideband pulses only act on one state of the superposition at a time. The angles $\theta,\varphi$ are controlled by the initial transmon $\ket{g}$-$\ket{e}$ pulse (indicated by an arrow), and the sequence following it realizes a unitary that encodes the qubit into a NOON state superposition for a cavity state starting off in the vacuum. 

We first prepare symmetric maximally entangled NOON states ($\theta = \pi/2$) and characterize them using photon-number-resolved spectroscopy of the transmon, shown in Figure~\ref{fig4}(b) (performed here without post-selection). The dispersive interaction leads to the cavity states $\ket{N0}$ and $\ket{0N}$ having peaks in the transmon spectrum that differ in frequency by $N\Delta\chi$, where $\Delta\chi$ is the difference in dispersive shifts between the two modes. The similar dispersive shifts for the modes of our system ($\Delta\chi \approx \qty{30}{\kilo\hertz}$) allow us to only resolve the peaks for higher photon number NOON states ($N>2$). However, the experiment demonstrates that the state does not have populations in other photon numbers to under $\qty{2}{\percent}$. To confirm the coherence of our state preparation, we perform a Ramsey-like experiment where we decode the NOON states $\ket{N0}$ and $\ket{0N}$ back to the transmon $\ket{g}$ and $\ket{e}$ states. We then apply a $\frac{\pi}{2}$-pulse with varying phase and observe oscillations that verify the coherence of the prepared NOON states. By varying the delay time between the encode and the decode operations, we can measure the decay time of the NOON state, finding it to be $1.27\pm \qty{0.05}{\milli\second}$ for $N=1$ (see Section S10 of the Supplementary Information). 

Additionally, we characterize NOON states up to $N=2$ using multimode Wigner tomography~\cite{chakram2020multimode, he2024efficient}. This is implemented through the pulse sequence shown in Figure~\ref{fig4}(d) (top) by displacing both cavities to a set of chosen displacements and measuring a generalized parity-like operator $\widetilde{\Pi} = \cos\left(\theta_1 \hat{n}_1+\theta_2 \hat{n}_2\right),$ where $\hat{n}_i$ is the photon number in mode $i$. The measurement of $\widetilde{\Pi}$ is performed through a transmon Ramsey experiment with delay time $\tau_R$, so that $\theta_i = \chi_i\tau_R \approx \pi$. The density matrix is then obtained by inverting the measurements while enforcing physicality constraints. The result for $N=2$ is shown in Figure~\ref{fig4}(d) (bottom). We extract post-selected fidelities of $\mathcal{F} = 0.95 \pm 0.012$  and $\mathcal{F} = 0.905 \pm 0.012$ for NOON states with $N= 1$ and $2$, respectively. Lastly, we verify that we can accurately prepare any state within the $\{\ket{N0}, \ket{0N}\}$ subspace by varying the amplitude and phase of the initial transmon pulse, decoding the state, and subsequently performing qubit tomography. We visualize the resulting states on the $\{\ket{N0}, \ket{0N}\}$ Bloch sphere, as shown in Figure~\ref{fig4}(e) for $N=3,4$. This confirms the expected trajectory and the ability to prepare the state at an arbitrary point in the encoded NOON state subspace. 

\begin{figure*}
  \begin{center}
    \includegraphics[width= 1.0\textwidth]{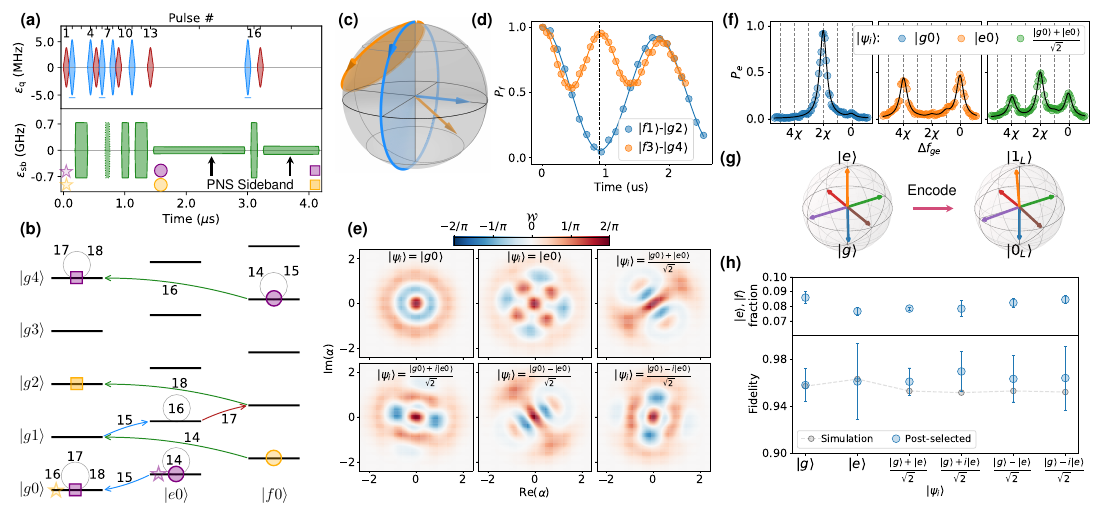}
    \caption{\textbf{Unitary encoding operation for the binomial code.} 
    \textbf{(a)} Pulse sequence for an encoding operation which maps transmon states to binomial code states ($\ket{g}\rightarrow\ket{2}$ and $\ket{e}\rightarrow(\ket{0}+\ket{4})/\sqrt{2}$). The rate-limiting photon-number-selective (PNS) sidebands are indicated by the arrows and the shelving pulses are indicated by underlines.
    \textbf{(b)} Energy level diagram of the combined Hilbert space of the transmon and a cavity mode. The gold circles represent the state after pulse 13 of the encoding sequence for the initial state of $\ket{g0}$ (gold star), and the purple circle corresponds to the initial state of $\ket{e0}$ (purple star). The subsequent pulses use PNS sidebands to bring the state back onto the $\ket{g}$ manifold of the transmon, completing the encoding gate. State transitions are tracked by arrows, while the gray circles denote no state change for that particular superposition branch during a given pulse. The final binomial code states are indicated by purple and orange squares. Numbers indicate the states occupied at the end of each corresponding pulse in the sequence. 
    \textbf{(c)} Bloch sphere illustration showing the principle of PNS sidebands, with a $\pi$ rotation applied to the target transition and cycling of the second transition.
    \textbf{(d)} Rabi oscillations for the $\ket{f1} \leftrightarrow \ket{g2}$ and $\ket{f3} \leftrightarrow \ket{g4}$ transitions, showing the timing for a $\pi$ and $2\pi$ rotation, respectively, corresponding to pulse 18 of the encoding sequence.
    \textbf{(e)} Reconstructed Wigner functions of the cavity states after the encoding operation, demonstrating the preparation of corresponding binomial code states when the transmon is initialized in six cardinal states. 
    \textbf{(f)} Photon-number-resolved spectroscopy of the prepared Binomial code states. \textbf{(g)} Bloch sphere representation of the encoding operation, illustrating the transition from the transmon Bloch sphere to the logical binomial code Bloch sphere. 
    \textbf{(h)} (Top) The fraction of the population not in $\ket{g}$ after the encoding operation for six cardinal transmon states. (Bottom)  Wigner tomography fidelities of the binomial code states following the encoding operation (blue markers) compared against the expected fidelities from Lindblad master equation simulations (gray markers).}
      \label{fig5}
  \end{center}
\end{figure*} 

The space of two-mode superposition states that can be realized with sideband transitions and transmon rotations, by shelving in the $|e\rangle$ manifold, is restricted to basis states of the form $\{\ket{n0}, \ket{0p}\}$. We introduce a second control technique that allows for the implementation of unitary operations that encode arbitrary multimode Fock states $\{\ket{n_1n_2\dots}, \ket{p_1p_2\dots}\}$, as shown in Section S4 of the Supplementary Information. This method addresses the challenge of realizing gates amid incommensurate sideband transition rates across different photon numbers by leveraging the residual dispersive interaction, which is not available on other hardware platforms, such as trapped ions and mechanical resonators. As shown in Figure~\ref{fig2}(b), the dispersive interaction causes sideband transitions corresponding to different photon numbers ($|f,n\rangle \rightarrow |g,n+1\rangle$) to vary in frequency by $n\chi_f$. These dispersive shifts allow us to use the detuning of the sideband drive as a control parameter to synchronize transitions for different photon numbers. This enables \textit{photon-number-selective} (PNS) sideband transitions, where we perform a $\pi$ pulse on a specific photon number while effectively performing no transition (e.g., $2\pi$ rotation) on another state. This technique is similar to those applied in other superconducting gate protocols~\cite{chapman2023high, tsunoda2024error}.

\section{Sideband-based encoding unitary for the binomial code}
We demonstrate our new method of unitary synthesis by implementing an encoding gate for the binomial code—an operation that maps an arbitrary qubit in the transmon $\ket{g}$-$\ket{e}$ subspace, $u\ket{g} + v\ket{e}$, onto a corresponding superposition of binomial code states, $u\ket{2} + v\frac{\ket{0}+\ket{4}}{\sqrt{2}}$. While this unitary has previously been demonstrated in the strong-dispersive regime~\cite{hu2019quantum, ni2023beating} (with $\chi\sim\qty{2}{\mega\hertz}$), we demonstrate a fast encoding gate in the weak-dispersive regime with nearly an order of magnitude smaller $\chi$ than in previous demonstrations. We implement our sideband-based binomial encoding unitary using the pulse sequence shown in Figure~\ref{fig5}(a). The sequence is again composed entirely of sideband and transmon rotations, with the key addition of two PNS sidebands, marked by the black arrows. The state transitions effected by the PNS sidebands (following pulse 13 of the sequence) are illustrated in the energy level diagram in Figure~\ref{fig5}(b). The complete sequence of state transitions for both $\ket{g0}$ and $\ket{e0}$ input states is provided in Fig. S15 of the Supplementary Information. 

In our scheme, the PNS sidebands are crucial for separating different cavity Fock states (e.g., $\ket{g2}$ and $\ket{g4}$) into distinct transmon manifolds, enabling these states to be addressed either simultaneously with transmon pulses or independently with fast, unselective sideband pulses through the use of shelving. Specifically, the two PNS sidebands perform a $\pi$-pulse for the $\ket{f1}\leftrightarrow\ket{g2}$ and $\ket{f0}\leftrightarrow\ket{g1}$ transitions, while implementing $2\pi$ and $4\pi$ rotations, respectively, for the $\ket{f3}\leftrightarrow\ket{g4}$ transition. These selective sidebands are the rate-limiting steps of the operation, scaling with the dispersive interaction but benefiting from favorable scaling factors due to their dependence on the dispersive shift of the $\ket{f}$ state and the involvement of higher photon numbers in the binomial code (see Section S9 of the Supplementary Information). With this, we implement the binomial encoding gate in a time shorter than the timescale set by the $\ket{e}$ dispersive shift, $1/\chi_e$, making the gate fast in an intermediate-$\chi$ regime where Purcell-limited cavity lifetimes can still be long. This gate speed can be further enhanced by using the transmon straddling regime~\cite{koch2007charge} to enhance $\chi_f$ without a commensurate increase in cavity Purcell decay. In our work, the inverse-Purcell-limited cavity lifetime is approximately \qty{60}{\milli\second}, which is comparable to the best reported lifetimes in niobium-based cavities~\cite{milul2023superconducting, romanenko2020three, oriani2024niobium}.

We characterize the unitary gate by performing Wigner tomography of the cavity states following the encoding operation, after initializing the transmon at six cardinal points on the transmon $\ket{g}$-$\ket{e}$ Bloch sphere, as shown in Figure~\ref{fig5}(e). The Wigner tomography is implemented with post-selection to characterize the cavity state accurately. The resulting states are consistent with expectations and are further verified by photon-number-resolved spectroscopy of the transmon, shown in Figure~\ref{fig5}(f), which displays peaks corresponding to the expected even Fock states, with populations consistent with those extracted from Wigner tomography to $\sim\qty{2}{\percent}$. We illustrate the encoded unitary by projecting the density matrices extracted from Wigner tomography into the subspace of binomial code states and reconstructing the states in the Bloch sphere of logical qubit, as shown in Figure~\ref{fig5}(g). The fraction of the shots with the transmon not in $\ket{g}$ range from $8$\textendash$\qty{9}{\percent}$ as shown in Figure~\ref{fig5}(h) (top). The post-selected fidelities (on $\ket{g}$) for the different fiducial states are shown in Figure~\ref{fig5}(h) (bottom) and range from $95.6$\textendash$\qty{96.9}{\percent}$. These results are consistent with master equation simulations (see Section S9 in the Supplementary Information) and are near the coherence limit. The detailed error budget for the Binomial encoding gate is shown in Fig. S17 of the Supplementary Information.  We find the gate infidelity to be dominated by ancilla decoherence, even for the post-selected fidelities. With state-of-the-art transmon coherence times ($T_1, T_{\varphi} = 500, \qty{200}{\micro\second}$), we expect post-selected $\ket{g}$ fractions of $\qty{0.3}{\percent}$ and binomial encode fidelities of $\geq \qty{99}{\percent}$ without post selection. The fidelities can also be improved by using mid-circuit measurements and feed-forward. These can be selectively implemented only for the slower PNS sidebands to significantly reduce infidelity.

\section{Conclusion}

In summary, we have used transmon-mediated charge-driven sideband interactions to demonstrate fast sideband control of a weakly coupled multimode bosonic memory. We achieve transmon-cavity SWAP gates in 10 modes in $150$\textendash$\qty{350}{\nano\second}$, which is $15$\textendash$30$ times faster than the bare dispersive coupling. Using this, we implement a tunable Jaynes--Cummings interaction within the transmon $\{\ket{g},\ket{f}\}$ subspace and introduce two strategies to address challenges associated with implementing unitary gates: a shelving technique that prevents unwanted transitions by temporarily storing populations in the sideband-transparent $\ket{e}$ manifold of the transmon, and exploiting the dispersive shift to synchronize sideband transition to implement photon-number-selective sideband gates. Using this, we demonstrated few-microsecond gates that encode a qubit into vacuum-Fock superposition in arbitrary modes and NOON states in arbitrary pairs of modes. Our encoding protocol is straightforwardly generalizable to arbitrary multimode Fock encodings. In addition, we have also prepared binomial code states across ten cavity modes. Finally, we implemented a new binomial encoding gate with average fidelities of $\qty{96.3}{\percent}$ for the encoded binomial states after post-selection. This gate is executed in approximately $\qty{4}{\micro\second}$, comparable with previous demonstrations, but with a bare dispersive coupling that is an order of magnitude weaker. 

The shelving techniques introduced in this work can be extended to implement a larger space of gates by using higher transmon states~\cite{wang2024systematic, rosenblum2018fault}. For example, access to the $\ket{h}$ state allows for autonomous error correction following a controlled-parity gate for the binomial code. The techniques presented in this work are highly relevant for implementing fast sideband-based gate operations in quantum processors and memories comprised of state-of-the-art ultra-low-loss niobium cavities without spoiling cavity coherence~\cite{LuAPS2025}. While building a multimode random-access memory requires fixing many-body coherent errors with new random-access architectures~\cite{RijuSchusterAPS2022}, the large contrast between the speed of sideband gates and the bare dispersive shift that we have demonstrated can potentially allow for their mitigation through frequency-robust sideband and transmon pulses~\cite{you2024crosstalk} and dynamical decoupling. Our control techniques allow for efficient control of multiple logical qubits in our multimode memory and tests of gate operations, code concatenation, and multimode bosonic codes~\cite{chuang1997bosonic, albert2019pair, Royer2022_Multimodegrid}. Our results provide a promising path toward fast bosonic unitary operations in multimode circuit QED systems, addressing key challenges in bosonic quantum error correction and quantum computing.

\section*{Acknowledgements}
We are grateful for insightful discussions with David Schuster, Michael Gershenson, Jens Koch, Tanay Roy, Xinyuan You, Taeyoon Kim, Shyam Shankar, Wolfgang Pfaff, Angela Kou, and Yuan Liu.  We acknowledge Cole Woloszyn and Paul Pickard of the Rutgers Physics and Astronomy Machine Shop for the cavity fabrication. We acknowledge David Schuster, Connie Miao, and Sara Sussman for their support in setting up the QICK measurement hardware; Taekwan Yoon and Zhixin Wang for the setup of Zurich instruments hardware; and David Van Zaanten, Alexander Romanenko, and Anna Grassellino for their support in transmon fabrication. We sincerely thank Michael Gershenson for providing access to his laboratory space and appreciate the support of Michael Gershenson, Plamen Kamenov, Ethan Kasaba, Andre Barbosa, and Raymond Niu in setting up the dilution refrigerator and microwave hardware used in these measurements. 

This work is supported by the U.S. Department of Energy, Office of Science and National Quantum Information Science Research Centers, Superconducting Quantum Materials and Systems Center (SQMS) under contract number DE-AC02-07CH11359, and by the Army Research Office under Grant Number W911NF-23-1-0096 and W911NF-23-1-0251.

\bibliography{references}

\end{document}


\title{Supplementary Information: Fast Sideband Control of a Weakly Coupled Multimode Bosonic Memory}

\author{Jordan Huang}
\email{jah499@scarletmail.rutgers.edu}
\affiliation{Department of Physics and Astronomy, Rutgers University, Piscataway, NJ 08854, USA}
\author{Thomas J. DiNapoli}
\affiliation{Department of Physics and Astronomy, Rutgers University, Piscataway, NJ 08854, USA}
\author{Gavin Rockwood}
\affiliation{Department of Physics and Astronomy, Rutgers University, Piscataway, NJ 08854, USA}
\author{Ming Yuan}
\affiliation{Pritzker School of Molecular Engineering, University of Chicago, Chicago, Illinois 60637, USA}
\author{Prathyankara Narasimhan}
\affiliation{Department of Physics and Astronomy, Rutgers University, Piscataway, NJ 08854, USA}
\author{Eesh Gupta}
\affiliation{Department of Applied Physics, Stanford University, Stanford, California 94305, USA}
\author{Mustafa Bal}
\affiliation{Superconducting Quantum Materials and Systems Division, Fermi National Accelerator Laboratory (FNAL), Batavia, IL 60510, USA}
\author{Francesco Crisa}
\affiliation{Superconducting Quantum Materials and Systems Division, Fermi National Accelerator Laboratory (FNAL), Batavia, IL 60510, USA}
\author{Sabrina Garattoni}
\affiliation{Superconducting Quantum Materials and Systems Division, Fermi National Accelerator Laboratory (FNAL), Batavia, IL 60510, USA}
\author{Yao Lu}
\affiliation{Superconducting Quantum Materials and Systems Division, Fermi National Accelerator Laboratory (FNAL), Batavia, IL 60510, USA}
\author{Liang Jiang}
\affiliation{Pritzker School of Molecular Engineering, University of Chicago, Chicago, Illinois 60637, USA}
\author{Srivatsan Chakram}
\email{schakram@physics.rutgers.edu}
\affiliation{Department of Physics and Astronomy, Rutgers University, Piscataway, NJ 08854, USA}
\date{\today}

\maketitle

\tableofcontents

\section{Multimode processor}
\subsection{Device design and fabrication}
The device consists of a multimode storage cavity and a readout cavity, bridged by a transmon. Both are seamless rectangular waveguide cavities fabricated using the flute method~\cite{chakram2020seamless}, created by drilling offset, overlapping, evanescent holes into a monolithic block of 5N5 aluminum. Microwave signals are applied through four SMA ports: a direct storage cavity drive, a direct transmon drive, and two readout cavity drives (input and output). Measurement information is extracted through the single output line connected to the readout cavity. The transmon is positioned to achieve relatively weak coupling to the storage modes, resulting in a dispersive shift of approximately 100-200 kHz.

After machining, the cavity surfaces undergo an etching process similar to that described in \cite{reagor2016superconducting}. The cavity is submerged in Transene Aluminum Etchant Type A, heated at \qtyrange{30}{50}{\degreeCelsius}, with constant agitation provided by a magnetic PTFE stirrer at 800 RPM. The process is carried out in two intervals: the first lasting two hours and the second lasting one hour. Between intervals, the used etchant is replaced to prevent saturation, and the cavity is immersed in a bath of deionized (DI) water. Following etching, the cavity is rinsed thoroughly in DI water and dried with nitrogen gas. This procedure removes approximately \qty{45}{\micro\meter} of material, as verified using a test block of 5N5 aluminum treated concurrently.

The transmon was fabricated by the Superconducting Quantum Materials and Systems (SQMS) center at Fermi National Laboratory at the Pritzker Nanofabrication Facility (PNF) at the University of Chicago, using the recipe outlined in~\cite{bal2024systematic}. The sputtered base layer consists of a Niobium underlayer capped with Tantalum. The Al-AlOx-Al Josephson junctions were fabricated by double-angle shadow evaporation using the Dolan bridge method.

\section{Quantum, microwave, and cryogenic hardware}

\begin{figure*}
    \includegraphics[width=1.\textwidth]{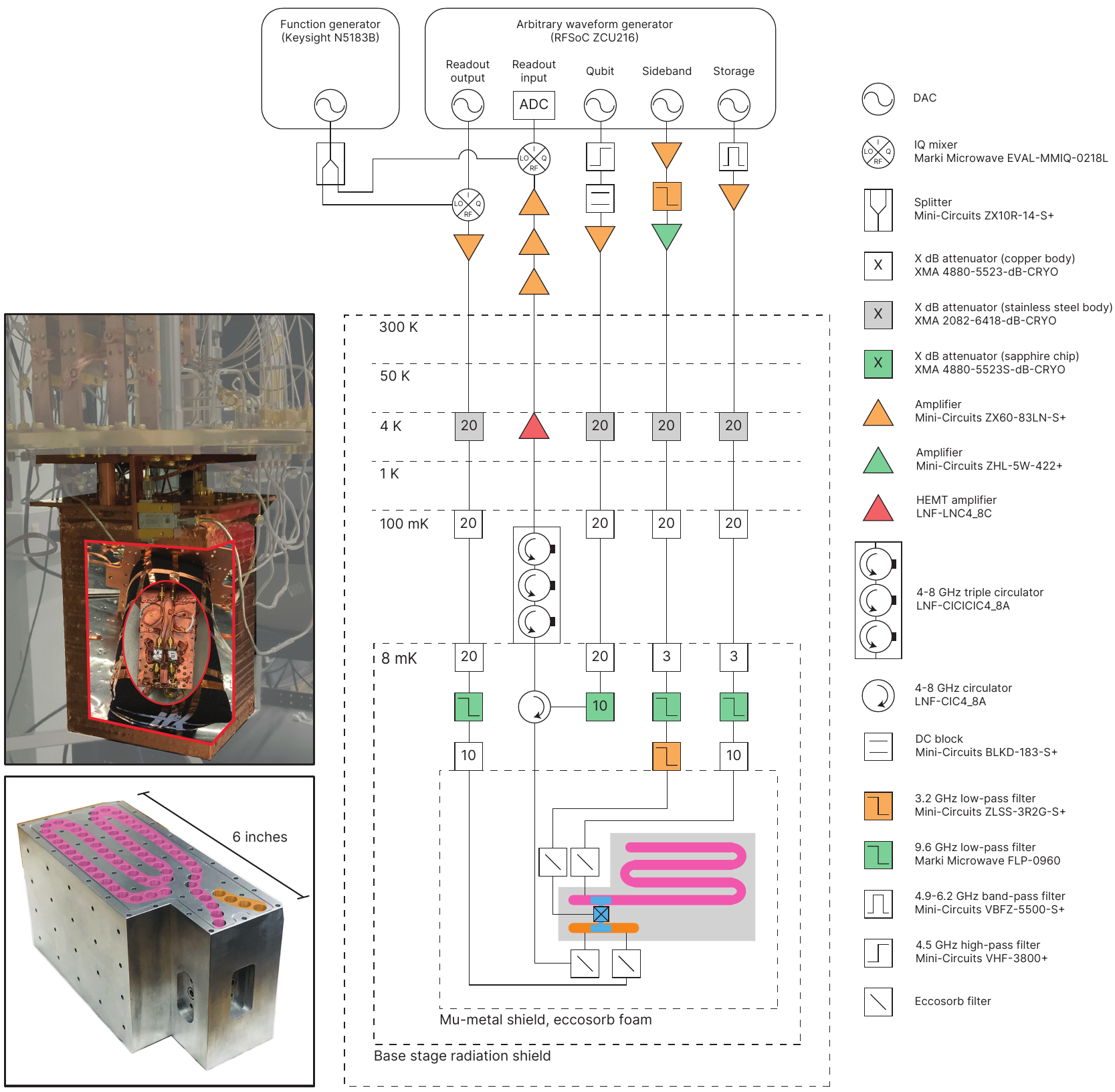}
    \caption{\textbf{Wiring diagram of the experimental setup.} A composite photograph (top left) shows the mu-metal can attached to the base stage of the dilution refrigerator, inside of which is the cavity, wrapped in eccosorb foam. The oval inset displays copper cavity-mounting plate, with readout cavity microwave lines and filters on the near side and the cavity bolted to the opposite side. A photograph (bottom left) shows the cavity device without accessories, with the storage cavity highlighted in pink and the readout cavity in orange. (Right) Wiring diagram for the microwave and cryogenic setup. }
    \label{fig:wiring_diagram}
\end{figure*}

For the bulk of the measurements presented in the paper, the control and readout pulses were digitally synthesized using a Xilinx ZCU216 RFSoC FPGA running the Quantum Instrumentation Control Kit (QICK)~\cite{stefanazzi2022qick} firmware and software. For the readout pulse, a signal from the RFSoC ZCU216 is mixed with an LO signal from a microwave function generator (Keysight N5183B). Figure~\ref{fig:wiring_diagram} displays a wiring diagram of the experimental setup.  Initial characterization of the cavity device was performed using the Zurich Instruments SHFQC Qubit Controller. The required phase stability between qubit, sideband, and cavity drive tones was achieved by taking the frequencies of the local oscillators associated with each drive to be integer multiples of the clock frequency. 

The cavity is bolted to a copper plate that is attached to the base stage ($\qty{8}{\milli\kelvin}$) of a Bluefors LD-400 dilution refrigerator. The cavity is enclosed inside a mu-metal can that is wrapped outside with copper tape. Surrounding the cavity inside the mu-metal can is eccosorb foam (Fig.~\ref{fig:wiring_diagram}).

\section{Multimode system Hamiltonian and Lindbladian}

\subsection{Multimode Jaynes-Cummings model with a transmon}

The Hamiltonian of our system is described by $\mathcal{H} = \mathcal{H}_{\text{sys}} + \mathcal{H}_\text{d}$ where
\begin{align}\label{sys_hamiltonian}
\begin{split}
    \mathcal{H}_{\text{sys}} &= \underbrace{\vphantom{\sum_i}\widetilde{\omega}_q c^\dagger c - E_J\left( \cos(\varphi) + \frac{\varphi^2}{2}\right)}_{\text{Transmon}} \\
    & \phantom{=} \ + \sum_i \biggl( \underbrace{\vphantom{\sum_i}\widetilde{\omega}_{c,i} b_i^\dag b_i}_{\text{Cavity}} +  \underbrace{\vphantom{\sum_i}g_i ( c - c^\dagger )(b_i-b_i^\dagger)}_\text{Coupling} \biggr).
\end{split}
\end{align}
Here, we have multiple cavity modes (with lowering operator $b_i$, where $i$ is the mode index) electric-dipole coupled to the same transmon and have written the transmon in the linearized oscillator basis, where $c$ is its lowering operator. $\widetilde{\omega}_q$ and $\widetilde{\omega}_{c, i}$ are the bare frequencies of the linear part of the transmon and cavity modes, respectively. $\varphi=\varphi_{\text{zpt}}(c+c^\dag)$ is the superconducting phase across the transmon's Josephson junction. Additionally, we have direct transmon and cavity drives given by 
\begin{equation}\label{eq:classical_drive}
\begin{split}
    \mathcal{H}_\text{d} &= -2i\epsilon_q \cos(\omega_{d} t + \phi_d)( c - c^\dagger ) \\
    & \phantom{=} \ -2i\sum_i \epsilon_{c,i}\cos(\omega_{d,c,i} t + \phi_{d, c, i})(b_i - b_i^\dagger).
\end{split}
\end{equation}
\subsection{Charge-driven sideband interactions}

We derive the sideband interaction using a four-wave mixing picture. We begin by considering the transmon coupled to one of the cavity modes, along with a direct transmon drive to activate the interaction. In order to prepare our Hamiltonian in an interaction frame, we first diagonalize the coupling between the linear parts of the system. Henceforth, $\omega_q$ and $\omega_{r}$ will refer to the dressed frequencies of the linear part of the transmon and cavity, respectively. Next, we go into the displaced frame of the drive with $U = \exp(\alpha^*(t) c - \alpha(t)c^\dagger$) where $\alpha(t) = e^{-i\omega_dt} \epsilon_q / (\omega_d - \omega_q) - e^{i \omega_d t} \epsilon_q / (\omega_d + \omega_q)$. Finally, we go into the rotating frame of the transmon and the cavity with $U = \exp\left(i\omega_q c^\dagger ct\right)\exp(i\omega_c b^\dagger bt)$. The phase operator transforms as
\begin{align}
\begin{split}
    \mathcal{\varphi} = \varphi_{\text{zpt}}\Bigl(& \beta_q e^{-i\omega_q t}c +  \beta_c e^{-i\omega_c t}b \\ 
    &+ \beta_qie^{-i\phi_d} e^{-i\omega_d t}\xi + \text{h.c.}\Bigr)
\end{split}
\end{align}
where
$\beta_q$ and $\beta_c$ are the participations of the transmon and the cavity in the phase across the junction, respectively. In the dispersive regime where $g \ll |\widetilde{\omega}_q - \widetilde{\omega}_r|$, we have $\beta_q \approx 1$ and $\beta_c \approx -g/\widetilde{\Delta}$ where $\widetilde{\Delta} = \widetilde{\omega}_q - \widetilde{\omega}_c$. $\varphi_\text{zpt} = \left( 2E_C/E_J\right)^{1/4}$ 
 is the zero-point fluctuation of the phase and $\xi=2\omega_d\epsilon_q/(\omega_d^2-\omega_q^2)$ is the effective transmon displacement. When $|\varphi| \ll 1$, we may expand the nonlinearity of our system as
\begin{equation}
    \mathcal{H_\text{nl}} = -E_J \left( \frac{\varphi^4}{4!} - \frac{\varphi^6}{6!} + \cdots \right).
\end{equation}
The nonlinearity in conjunction with the transmon drive can activate parametric interactions which can be made resonant through a judicious choice of the drive frequency $\omega_d$. The fastest interactions involving the cavity mode use the fourth-order nonlinearity $\varphi^4$ to convert two transmon photons into a photon in any of the cavity modes. This gives rise to an interaction of the form $\widetilde{g}_{\text{sb}}(ie^{-i\phi_d}c^\dagger c^\dagger b -  ie^{i\phi_d}ccb^\dagger)$, where
\begin{equation}\label{eq:g_sb}
    \widetilde{g}_\text{sb} = \frac{gK\xi}{\Delta}.
\end{equation}
Here, $K \approx -E_C$ is the transmon anharmonicity and $\Delta=\omega_q - \omega_c$, where we have made the approximation $\Delta \approx \widetilde{\Delta}$. This interaction can connect any two transmon levels that differ by two-photons, such as $\ket{f,n}$ and $\ket{g,n+1}$ or $\ket{h,n}$ and $\ket{e,n+1}$. The $\ket{f,n}$-$\ket{g,n+1}$ sideband interaction is made resonant by choosing $\omega_d = 2\omega_q + K - \omega_c$, the difference frequency between the transmon's $\ket{f}$ state and the cavity. A SWAP operation in the $\ket{g}$-$\ket{f}$ manifold is given by $g_{\text{sb}}(ie^{-i\phi_d}\ket{f}\bra{g} b -  ie^{i\phi_d}\ket{g}\bra{f}b^\dagger)$, where
\begin{equation}\label{eq:g_sb_f0g1}
   g_\text{sb} = \frac{\sqrt{2}gK\xi}{\Delta} = \frac{\sqrt{2}(\omega_q + K +\Delta)}{2\omega_q + K + \Delta}\frac{\epsilon}{g}\chi.
\end{equation}
Here, $\chi=2g^2K/(\Delta(\Delta+K))$ is the dispersive shift of the $\ket{e}$ state. In this form, we see that $\epsilon$ can be increased such that this interaction strength is larger than $\chi$. Including the contribution from higher-order nonlinearities (beyond $\varphi^4$), while making an RWA to drop counter-rotating interaction terms, modifies the sideband rate to 
\begin{equation}
    \widetilde{g}_\text{sb} = \frac{gK\xi}{\Delta}\sum_{n=1}^\infty \frac{(-1)^{n+1}(\varphi_\text{zpt}\xi)^{2n-2}}{(n-1)!(n-1)!}.
\end{equation}
We see that the sideband rate becomes nonlinear in drive strength $\epsilon$. Because of the opposite signs of successive nonlinear terms, this formula predicts a downturn in the rate at sufficiently large $\epsilon$.

\subsection{Calibrating the dispersive and Kerr shifts}

In the dispersive regime where $g \ll \left|\omega_q - \omega_{c, i}\right|$, the system can be described by a simplified Hamiltonian
\begin{align}
\begin{split}
    \mathcal{H} &= \omega_q c^\dagger c + \frac{K}{2}c^\dagger c^\dagger cc \\ & \phantom{=} \ + \sum_i \biggl( \omega_{c, i} b^\dag_{i} b_i + \frac{K_i}{2} b^\dag_{i} b^\dag_{i} b_i b_i + \chi_i c^\dag c b^\dag_i b_i \biggr),
\end{split}
\end{align}
which can be derived from Eq. \eqref{sys_hamiltonian} using second-order perturbation theory. We measure $\omega_q$, the transmon's anharmonicity $K$ using standard $\ket{g}$-$\ket{e}$ and $\ket{e}$-$\ket{f}$ two-tone spectroscopy measurements that are refined with  transmon Ramsey measurements. We also measure the transmon life times using standard $T_1$ measurements, and extract decoherence times $T_2$ and $T_2^*$ using Ramsey experiments with and without a spin-echo pulse. We measure the transmon's thermal population ($\ket{e}$ population) by measuring the $\ket{f}$-state population after driving the $\ket{e}$-$\ket{f}$ transition, with and without an initial $\pi_{ge}$ pulse. 

We measure the cavity mode frequencies $\omega_{c, i}$ in the single-photon regime using the transmon via their cross-Kerr interaction (the dispersive shift $\chi_i$). This interaction changes the transmon frequency based on the photon number in the cavity. We perform cavity spectroscopy using a cavity drive pulse sent through the direct storage drive port while sweeping its frequency. The cavity state is subsequently probed by a narrow-band photon number-resolved transmon pulse driven on the zero-photon peak, which affects a $\pi$ pulse conditioned on the cavity being in vacuum, thereby measuring the zero-photon population. When on resonance, the cavity pulse excites the mode and reduces the zero-photon population, leading to a dip in the spectrum. The cavity mode frequency is further refined by a cavity-Ramsey experiment \cite{chakram2020multimode}, where we excite the cavity with two displacement pulse, while varying the time $\tau$ between them and advancing the phase of the second displacement pulse by $\omega_{\text{R}}\tau$). This is followed by a measurement of the zero-photon population through a resolved transmon $\pi$-pulse.

We measure the dispersive shift $\chi_i$ of the cavity modes by performing a transmon $\ket{g}$-$\ket{e}$ Ramsey experiment after adding a photon in the target mode using charge-sidebands, using the sequence $\pi_{ge} \rightarrow \pi_{ef} \rightarrow \pi_{f0g1}$. We also measure $\chi_{f,i}$ using the same protocol, with the Ramsey experiment between $\ket{g}$ and $\ket{f}$ transmon levels. With calibrated transmon pulses, the difference in the oscillation frequency from the Ramsey frequency programmed in software gives a precise measurement of $\chi_i$ limited only by linewidth of the transmon.

\subsection{Long-lived mode excited by readout pulse}\label{sec:spurious_mode}
\begin{figure}[t]
    \centering
    \includegraphics[width=0.5\textwidth]{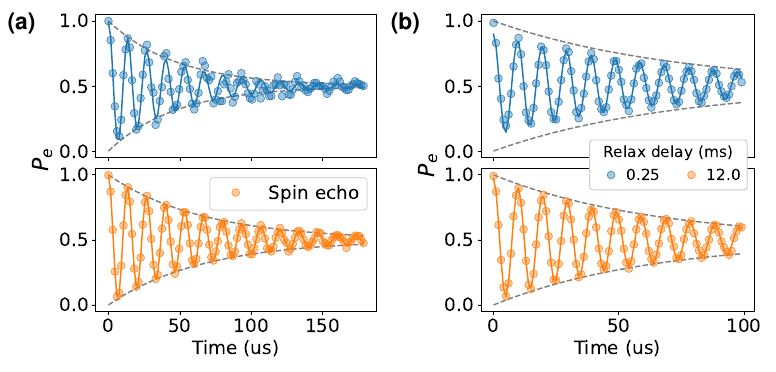} 
    \caption{\textbf{Spurious mode excited by readout pulse.} \textbf{(a)} Ramsey experiment with (top) and without (bottom) a spin echo. \textbf{(b)} Ramsey experiment with a $\qty{0.25}{\milli\second}$ delay time (top) and a $\qty{12}{\milli\second}$ delay time (bottom) between experiments for the same readout pulse settings.}
    \label{fig:spurious_mode}
\end{figure}
We observe spurious beating and reduced contrast in transmon Ramsey measurements, which depend on the magnitude and duration of the readout pulse at $\qty{8.017}{\giga\hertz}$, as well as the experiment duty cycle. These effects are eliminated by a spin-echo sequence, as shown in Fig.\ref{fig:spurious_mode}(a), and are also mitigated in a standard Ramsey experiment by reducing the duty cycle, as shown in Fig.\ref{fig:spurious_mode}(b). This suggests that the readout tone excites a long-lived mode in the system that is dispersively coupled to the transmon. Qubit spectroscopy indicates that this is a long-lived mode rather than a two-level system (TLS)~\cite{PhysRevLett.133.160602}, though its dispersive shift does not correspond to the storage cavity mode closest to the readout mode at $7.967$ GHz. We estimate the lifetime of the spurious mode to be on the order of milliseconds. To mitigate this issue, we minimize the readout drive power while maintaining readout fidelity and increase the time delay between experiments.

\begin{table*}
    \begin{tabular}{ l  c  c }
      \hline
      Parameter & Hamiltonian Term ($\hbar = 1$) & Value \\ 
      \hline
      Transmon frequency $\nu_{ge}$ & $\pi \nu_{ge}(\ket{e}\bra{e}-\ket{g}\bra{g})$ & $\qty{4.606}{\giga\hertz}$ \\

      Transmon frequency $\nu_{ef}$ & $\pi \nu_{ef}(\ket{f}\bra{f}-\ket{e}\bra{e})$ & $\qty{4.492}{\giga\hertz}$ \\

      Transmon frequency $\nu_{fh}$ & $\pi \nu_{fh}(\ket{h}\bra{h}-\ket{f}\bra{f})$ & $\qty{4.371}{\giga\hertz}$ \\
     
      Transmon $\ket{e}\rightarrow\ket{g}$ decay time $T_1$ & & $\qty{55.83}{\micro\second}$  \\

      Transmon $\ket{f}\rightarrow\ket{e}$ decay time $T_1$ & & $\qty{28.85}{\micro\second}$  \\
     
      Transmon $\ket{g}\text{-}\ket{e}$ dephasing time $T_2^*, T_2$ & & 47.20, 65.65  \qty{}{\micro\second} \\

      Transmon $\ket{g}\text{-}\ket{f}$ dephasing time $T_2^*, T_2$ & & 36.58, 41.06 \qty{}{\micro\second} \\
      
      Transmon thermal population $\bar{n}$ & & \begin{tabular}{@{}c@{}}$\qty{0.47}{\percent}$ \\ $\qty{0.32}{\percent}$ (with transmon reset) \end{tabular} \\
      \hline
      Storage mode frequency $\nu_{c}$ & $2\pi \nu_{c} b^\dag b$ & \begin{tabular}{@{}l@{}}$\{5.750, 5.994, 6.228, 6.479, 6.720, 6.962, $ \\ $ 7.216, 7.461, 7.715, 7.967\}\ \qty{}{\giga\hertz}$ \end{tabular} \\

      Storage mode dispersive shift $\chi$ & $2\pi\chi_e b^\dagger b \ket{e}\bra{e}$ & \begin{tabular}{@{}l@{}}$\{-197, -217, -202, -208, -171, -150,  $ \\ $ -165, -133, -117, -106\}\  \qty{}{\kilo\hertz}$ \end{tabular}\\
      
      & $2\pi\chi_f b^\dagger b \ket{f}\bra{f}$ & \begin{tabular}{@{}l@{}}$\{-356, -383, -394, -391, -363, -299, $ \\ $ -304, -245, -220, -190\}\  \qty{}{\kilo\hertz}$ \end{tabular}\\
      
      Storage mode single-photon decay time $T_1$ & &  \begin{tabular}{@{}l@{}}$\{1.187, 1.253, 1.298, 1.175, 1.175, 0.899, $\\ $0.656, 0.851, 0.936, 0.989\}\ \qty{}{\milli\second}$ \end{tabular} \\
      
      Storage mode single-photon dephasing time $T_2^{*}$ & & \begin{tabular}{@{}l@{}}$\{1.932, 2.044 , 2.0243, 1.926, 1.897, 1.559, $\\ $1.1613, 1.555, 1.662, 1.811\}\ \qty{}{\milli\second}$ \end{tabular} \\
      
      Storage mode thermal population $\bar{n}$ & & \begin{tabular}{@{}l@{}}$\{0.82, 0.85, 0.29, 0.54, 0.38, 1.30, 1.93\}\ \qty{}{\percent}$ \\ $\{0.26, 0.55, 0.48, 0.58, 0.49, 0.39, 0.59,$ \\ $0.62, 0.73, 0.70\}\ \qty{}{\percent}$ (with cavity reset) 
      \end{tabular}\\

      \hline
      Readout mode frequency $\nu_r$ & $2\pi \nu_r b^\dag_r b_r$ & $8.0174$ GHz \\
      Readout mode dispersive shift $\chi_e$ & $2\pi\chi_e b_r^\dagger b_r \ket{e}\bra{e}$ & $\qty{-845}{\kilo\hertz}$ \\
      Readout mode single-photon decay time $T_1$ & & $\qty{0.61}{\micro\second}$ \\
      Readout mode thermal population $\bar{n}$ & & \qty{0.54}{\percent} \\
      \hline
    \end{tabular}
    \caption{\textbf{Measured system parameters.}}
    \label{tab:system_params}
\end{table*}

\begin{figure*}
    \centering
    \includegraphics[width=1\linewidth]{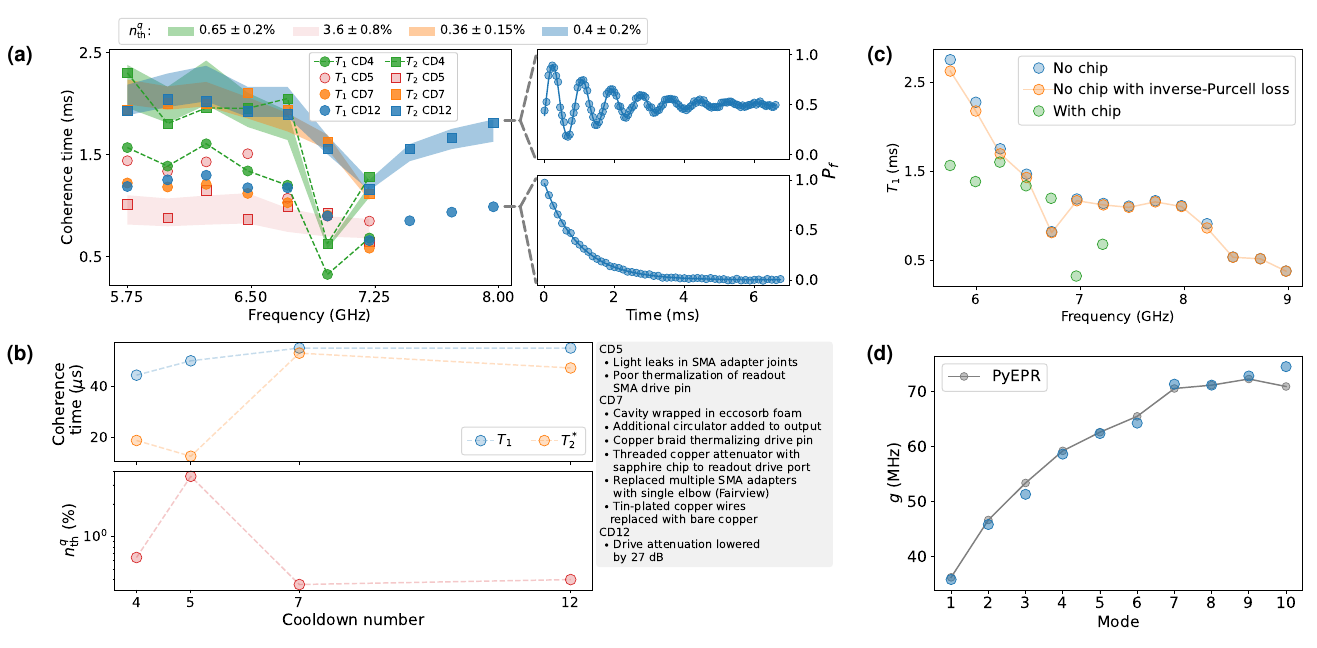}
    \caption{\textbf{Transmon-cavity characterization.} \textbf{(a)} Coherence times of the cavity modes across different cooldowns. The shaded regions indicate the inferred qubit thermal population that would give rise to the measured cavity $T_2$. The right inset shows the cavity $T_2$ Ramsey (top) and the $T_1$ experiment (bottom) for a given mode. \textbf{(b)} Transmon coherence times (top) and inferred transmon thermal population from cavity coherence (bottom) across different cooldowns. The grey box highlights important changes made between cooldowns. The major change that lowered the transmon thermal population was wrapping the cavity in eccosorb foam. The transmon $T_2^*$ was increased by a factor of $4$ through changes made to the output lines.  \textbf{(c)} Cavity $T_1$ with and without the transmon and sapphire substrate. \textbf{(d)} The blue circles denote the bare coupling ($g$) derived from experimentally measured $\chi$'s for the modes. The grey circles are predictions from pyEPR \cite{minevEnergyparticipationQuantizationJosephson2021} simulations.}
    \label{fig:system_characterization}
\end{figure*}

\subsection{Cavity mode coherence times}

We measure the single-photon lifetime ($T_1$) of a cavity mode by first adding a photon to the mode via the sequence $\pi_{ge} \rightarrow \pi_{ef} \rightarrow \pi_{f0g1}$. After waiting for a variable delay, we transfer the remaining one-photon population back to the transmon using a $\pi_{f0g1}$ pulse and measure the transmon’s $\ket{f}$ population as a function of wait time to extract the cavity’s $T_1$. To measure the single-photon dephasing time, we perform a cavity Ramsey experiment by preparing the superposition state $\frac{\ket{0}+\ket{1}}{\sqrt{2}}$ in the target mode. After allowing the state to idle in the cavity for a variable delay, we map the cavity states $\{\ket{0}, \ket{1}\}$ onto the transmon states $\{\ket{g}, \ket{e}\}$ using the pulse sequence $\pi_{f0g1} \rightarrow \pi_{ef}$, where we sweep the phase of the $\pi_{ef}$ pulse as a function of the delay time. Finally, we apply a transmon $\frac{\pi}{2}_{ge}$ pulse to convert the phase information into population, thereby performing a Ramsey experiment on the $\{ \ket{0}, \ket{1} \}$ subspace.

The resulting $T_1, T_2$ times for the cavity modes are shown in Fig. 1 of the main text and in Table \ref{tab:system_params}. We note that the cavity mode $T_1$ times were found to be lower than for the bare cavity without the transmon chip (Fig. \ref{fig:system_characterization}(c)). This additional decay was beyond that expected from the inverse-Purcell loss from transmon decay, estimated from the measured transmon-mode couplings. The extra loss is attributed to the sapphire substrate on which the transmon is patterned. We also note that the mode $T_1$'s degraded from the initial cooldowns (Fig. \ref{fig:system_characterization}(a)) despite no change made to the packaging and filtering. No change was observed in the transmon lifetimes across cooldowns.

A summary of the changes made between cooldowns and their effect on transmon coherences times and thermal populations is shown in Fig. \ref{fig:system_characterization}. The thermal population of the transmon was initially high, which we attributed to inadequate shielding from high-frequency infrared radiation above the superconducting gap, known to create quasiparticles. This was mitigated by wrapping the cavity in eccosorb foam, which reduced the thermal population from $3\% \rightarrow 0.5\%$.  The readout mode's thermal population was also reduced from $\qty{1.32}{\percent}$ to $\qty{0.54}{\percent}$ during the same cooldown. While there were several changes, we particularly note the additional circulator added to the readout output line, which likely reduced thermal photons leaking in through it. 

\subsection{Cavity mode thermal occupations}

We measure the cavity thermal population by applying an $\ket{f0}$-$\ket{g1}$-pulse of varying duration on the target cavity mode with the transmon initialized in the ground state. Given the negligible thermal population of the transmon $\ket{f}$ state, the small residual oscillation that is observed is due to the thermal occupation of the one-photon state of the mode. We measure this population precisely by comparing the oscillation with that observed with the transmon prepared in $\ket{f}$. The resulting thermal occupations are show in Fig. \ref{fig:sequential_sideband_reset}(d) ranging from $0.2$--$2\%$ across the modes. We note that the modes that have higher thermal occupation also have lower quality factor from coupling to the direct storage drive line. In order to mitigate this state preparation error, we perform a sequential sideband reset protocol prior to the start of every experiment, where we transfer the cavity population to the $\ket{f}$ and cool the state by driving the $\ket{f0_r}$-$\ket{g1_r}$ transition with the readout cavity, which has a $T_1$ of \qty{0.61}{\micro\second} (see Section \ref{sec:sequential_sideband_reset}). 

\begin{figure}
    \centering
    \includegraphics[width=\linewidth]{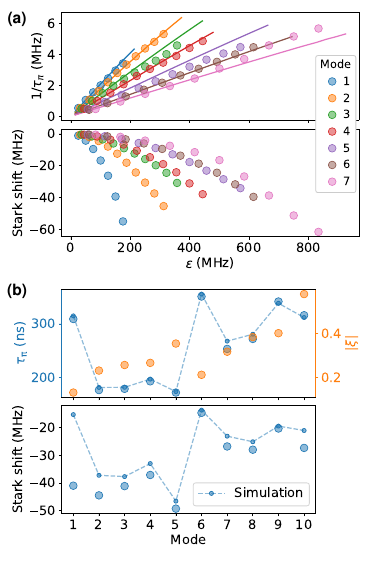}
    \caption{\textbf{Sideband characterization.} \textbf{(a)} (Top panel) Measured sideband rates (circles) as a function of transmon drive strength for the first seven modes of the storage cavity. Solid lines represent simulated rates obtained from Floquet numerics. (Bottom panel) Drive-induced Stark shift of the sideband resonance as a function of drive strength. \textbf{(b)} (Top panel) Measured $\pi$-pulse times (blue circles) for the $\ket{f0}$–$\ket{g1}$ sideband across modes at a fixed drive strength, corresponding to the one used for binomial state preparation experiments in Fig. 3(f) of the main text. Simulated $\pi$-pulse times (blue dots) and corresponding transmon displacement required to achieve the same speed (orange circles, right axis) are also shown. (Bottom panel) Measured drive-induced Stark shift of the sideband resonance (blue circles) at the corresponding $\ket{f0}$–$\ket{g1}$ drive strengths. Numerical predictions are shown as blue dots.}
    \label{fig:sideband_char}
\end{figure}

\section{Multimode cavity control with sidebands}

The sideband interaction allows us to engineer an effective multimode Jaynes-Cummings-like Hamiltonian where we have control of the coupling strength, given by

\begin{align}\label{eq:tunable_jaynes_cummings}
\begin{split}
    \mathcal{H} = \sum_i \biggl(& \underbrace{\vphantom{\sum_i} \chi_{e,i} b^\dag_i b_i \ket{e}\bra{e} + \chi_{f,i} b^\dag_i b_i \ket{f}\bra{f}}_\text{Dispersive shift} \\ & + \underbrace{\vphantom{\sum_i}g_{\text{sb}, i}(t) (b_i\ket{f}\bra{g} + b_i^\dag \ket{g}\bra{f})}_{\text{Tunable coupling}}\biggr)
\end{split}
\end{align}

The tunable Jaynes-Cummings coupling, combined with transmon drives on the $\ket{g}$-$\ket{e}$, $\ket{e}$-$\ket{f}$, and direct $\ket{g}$-$\ket{f}$ two-photon Raman transition, provides a versatile set of interactions that enable the intuitive derivation of analytic pulse sequences for gate operations using a Jaynes-Cummings ladder picture.

A key challenge when relying solely on sideband interactions ($g_{\text{sb}} \gg |\chi|$) is their lack of photon-number selectivity and the incommensurate sideband Rabi rotation rates across different photon-number states. As detailed in the main text, we address this issue through two complementary strategies. First, we implement a shelving technique, where we temporarily store selected states in the transmon’s $\ket{e}$ manifold. Since the sideband interaction occurs only in the $\ket{g}$-$\ket{f}$ manifold and is detuned from the $\ket{e}$ manifold by twice the transmon anharmonicity, this suppresses transitions of the shelved states by a factor of $\sim10^{-4}$. Second, we leverage the dispersive interaction available in our system to introduce a photon-number-selective sideband pulse (see Section \ref{sec:pnr_sideband}), which implements selective sideband operations on a given photon number state while leaving another non-shelved state unchanged (up to a phase). 

\subsection{Universal encoding of multimode Fock states}\label{sec:multimode_fock_encoding}

We introduce a protocol to implement a unitary gate $U$ that encodes the transmon ground and excited states into two arbitrary multimode Fock states:
\begin{align}\label{eq:encoding_gate}
\begin{split}
     & U\left(\cos{\theta}\ket{g}+e^{i\varphi}\sin{\theta}\ket{e}\right)\otimes\ket{0\ldots0}\\ &=\ket{g}\otimes\left(\cos{\theta}\ket{n_1\ldots n_M}+e^{i\varphi}\sin{\theta}\ket{p_1\ldots p_M}\right).
\end{split}
\end{align}
We prove that we can construct this unitary using the following protocol. We first use sideband and qubit rotations to take $\ket{g}$ to a desired multimode Fock state ($\ket{g}\otimes\ket{n_1\ldots n_M}$) while $\ket{e}$ is left shelved ($\ket{e}\otimes\ket{0\ldots 0}$). Next, we periodically shelve the state $\ket{g}\otimes\ket{n_1\ldots n_M}$ while preparing the second multimode Fock state one photon at a time using transmon rotations and sideband transitions, up to a state one-photon away from the target ($\ket{e}\otimes\ket{0\ldots 0} \rightarrow \ket{f}\otimes\ket{p_1-1\ldots p_M} $). During the preparation of the second state, periodic shelving can be skipped if a direct $\ket{g}$-$\ket{f}$ transition is used. Before the final sideband pulse, we unshelve the first state into the $\ket{g}$-manifold ($\ket{g}\otimes\ket{n_1\ldots n_M}$). We then bring the second state to the $\ket{g}$ manifold using a photon-number-selective sideband pulse. This construction is not unique, as we can permute the order in which we climb the multimode Jaynes-Cummings ladder.

With this protocol, the number of sideband and qubit pulses required for each is $N_\text{tot}=\sum_{i=1}^{M}\left(n_i+p_i\right)$, in addition to shelving pulses. Crucially, the speed of the photon number-selective pulse is proportional to the detuning of the two encoding states with respect to their dispersive shifts (see Section \ref{sec:pnr_sideband}).  In the special condition where at least one of $n_i$ or $p_i$ are equal 0, a photon number-selective sideband is not required at the last step. The NOON states $(\ket{N0}+\ket{0N})/\sqrt{2}$ are an example of such a state, for which we implement an encoding gate with just fast sideband transitions (see Section \ref{sec:noon_state}).

\subsection{Comparison with Law-Eberly state preparation}
The control Hamiltonian in Eq. \ref{eq:tunable_jaynes_cummings} can be used to prepare arbitrary cavity states in a single mode using the Law-Eberly protocol~\cite{laweberly}. This protocol prepares an arbitrary quantum state by finding the sequence of sideband and qubit rotations that removes one photon at a time from the cavity mode to transition from the initial state to $\ket{g0}$. To prepare the target state, we simply apply the reverse of this gate sequence. 

\begin{figure}
    \centering
    \includegraphics[width=0.4\textwidth]{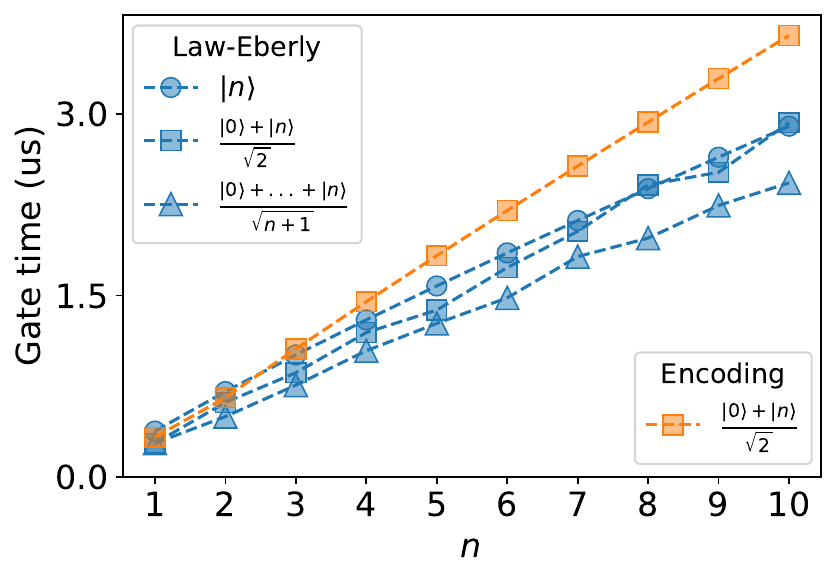} 
    \caption{\textbf{Comparison of the scaling with photon number for Law-Eberly state preparation and our encoding protocol.} The blue markers show state preparation times using the Law-Eberly protocol for various cavity states as a function of $n$. The orange markers show the gate times for encoding vacuum-Fock state superpositions as a function of $n$ using the scheme detailed in Section \ref{sec:multimode_fock_encoding}.}
    \label{fig:law_eberly_encoding_scaling}
\end{figure}
In contrast to~\cite{laweberly}, the Hamiltonian in Eq.\ref{eq:tunable_jaynes_cummings} also includes photon-number-dependent dispersive shifts in the qubit and cavity-mode frequencies. In the single-mode case, the original Law-Eberly protocol can be extended to incorporate the dispersive Hamiltonian in Eq.\ref{eq:tunable_jaynes_cummings} by introducing photon-number-dependent shifts in the sideband and qubit pulse frequencies.

With this protocol, the number of sideband and qubit pulses required each to implement the unitary $U$ that prepares an arbitrary state,
\begin{equation}
    U\ket{0}=\sum_{n=0}^{N}c_{n}\ket{n},
\end{equation}
is $N_\text{max}=N$, the highest number of photons in the state. Additionally, it is faster to prepare states whose populations are more distributed in the photon number basis (Fig. \ref{fig:law_eberly_encoding_scaling}). This speedup is because the qubit and sideband pulses that recombine superpositions in the photon number basis do not necessarily need to complete full $\pi$ rotations.

As shown in Fig.~\ref{fig:law_eberly_encoding_scaling}, the number of gates required for the encoding protocol scales linearly with the total number of photons in the superposition state. In contrast, the Law-Eberly state preparation protocol scales linearly with only the highest photon number in the target state. Importantly, we emphasize that the encoding protocol is not merely a state preparation method but a full gate operation that encodes transmon states into arbitrary multimode Fock states.

\section{Transmon and cavity reset with sequential sidebands}\label{sec:sequential_sideband_reset}

\begin{figure*}
    \centering
    \includegraphics[width=\linewidth]{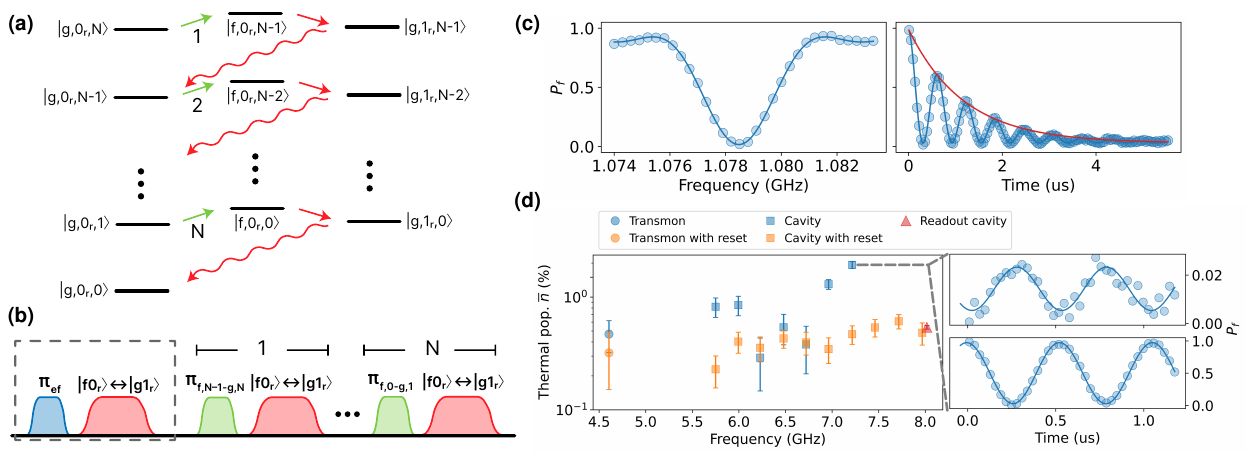}
    \caption{\textbf{Cavity and transmon reset using sequential sidebands.} \textbf{(a)} Energy level diagram of the sequential sideband reset protocol used to cool the $N$ photon state $\ket{gN}$. \textbf{(b)} Corresponding pulse diagram. The grey box indicates the pulse sequence to be played to cool the state $\ket{eN}$. \textbf{(c)} $\ket{f0}$-$\ket{g1}$ spectroscopy for the readout mode (left) and corresponding sideband Rabi oscillation (right) with a decay time of $600~\mathrm{ns}$. In the reset protocol, a storage mode sideband transfers photons into the transmon which is cooled through a sideband with the readout mode. \textbf{(d)} Transmon and cavity thermal populations before and after the reset protocol. Inset: $\ket{f0}$-$\ket{g1}$ Rabi oscillations for the seventh mode with (bottom) and without (top) preparing a photon in the cavity.}
    \label{fig:sequential_sideband_reset}
\end{figure*}

We increase the experiment's duty cycle by implementing sequential transmon and cavity reset, cooling both the transmon and cavity mode one photon at a time using sideband pulses between the transmon and the lossy readout cavity. Cooling the state $\ket{g,N}$ by one photon is achieved via a $\pi_{f,N-1\text{-}g,N}$ pulse, followed by driving the $\ket{f,0_r,N-1}$-$\ket{g,1_r,N-1}$ transition through the readout mode to dissipate the transmon excitation. This process is repeated $N$ times until the cavity reaches the ground state. Similarly, the $\ket{e,N}$ state is reset by first applying a $\pi_{ef}$ pulse, followed by driving the $\ket{f,0_r,N}$-$\ket{g,1_r,N}$ transition via the readout mode before proceeding with the same protocol used to cool $\ket{g,N}$.

These protocols and their corresponding pulse sequences are illustrated in Fig.~\ref{fig:sequential_sideband_reset}(a) and (b). A full reset cycle sequentially cools all $n$-photon states up to $N$ to the ground state for both $\ket{g}$ and $\ket{e}$ transmon states. Multiple reset cycles are applied to mitigate population redistribution caused by incommensurate rotations from fast sideband pulses and to remove population from cavity states where the transmon remains in $\ket{f}$. Spectroscopy and Rabi oscillations of the $\ket{f,0_r}$-$\ket{g,1_r}$ sideband with the readout mode is shown in Fig.~\ref{fig:sequential_sideband_reset}(c) indicating a readout mode decay time of \qty{0.61}{\micro\second}. The improvement in the transmon and cavity mode thermal populations with reset are shown in Fig.~\ref{fig:sequential_sideband_reset}(d). The final thermal populations following reset are comparable to the thermal population of the readout cavity, inferred independently from shot-noise dephasing of the transmon~\cite{clerk2007using}.

\section{Calibrating and optimizing sideband pulses}
Maximizing the fidelity of strongly driven interactions requires smooth pulses that allow adiabatic evolution of the undriven eigenstates to their corresponding Floquet modes (See Section \ref{sec:floquet_ramp}). As a result, we cannot simply use a rectangular pulse and must implement smooth ramps. Our sideband pulse shape is a flat top, ramped up and down using a bump function, with the pulse envelope given by,
\begin{equation}\label{eq:ramp_profile}
    \epsilon(t) = 
    \begin{cases}
        \epsilon_{\text{max}}\exp{(2+\frac{2}{\left(\frac{t-\tau}{\tau}\right)^2-1})},&  t \leq \tau \\
        \epsilon_{\text{max}},& \tau < t < T - \tau \\
        \epsilon_{\text{max}}\exp{(2+\frac{2}{\left(\frac{t-T+\tau}{\tau}\right)^2-1})},& t \geq T - \tau
    \end{cases}
\end{equation}
where $\tau$ is the ramp time and $T$ is the total length of the pulse. This function has the feature of being infinitely differentiable, with the ability to do fine adjustments of the effective pulse time. We note that we observed similar performance with a simpler $\mathrm{sin}^{2}$ ramp function. The envelope is digitally combined in the FPGA with a carrier at the sideband frequency $\omega_{\text{sb}}$.

The sideband interaction is induced by applying charge drives to the transmon, which leads to a drive-amplitude-dependent Stark shift of the transmon energies, which shifts the sideband resonance frequency. We prepare the transmon in $\ket{f}$ and find the resonance frequency through sideband spectroscopy. We sweep  $\omega_\text{sb}$ and measure the $\ket{f}$ population, fitting the resulting spectrum to a since function tp extract the resonance frequency. We subsequently drive the sideband on resonance and measure the $\ket{f}$ population as a function of pulse time to extract the $\pi$-time. We perform this calibration iteratively for different $\epsilon$, operating at an $\epsilon_{max}$ where there is no drop in sideband Rabi oscillation contrast.

We estimate the sideband interaction fidelity by fitting the Rabi oscillations to 
\begin{equation}\label{eq:rabi_fidelity}
    P_f = \frac{1}{2}e^{-\kappa t}(1+e^{-\kappa_\varphi t}\cos{(2g_{\text{sb}}t)}).
\end{equation}
Following \cite{lu2023high}, the fidelity of a $\pi$-pulse is given by $F\approx1-\frac{\pi}{2}\frac{\kappa+\kappa_\varphi/2}{g_\text{sb}}$, where $\kappa$ is the effective decay and $\kappa_\varphi$ the effective dephasing of the interaction. This inferred fidelity is plotted in Fig. 2(f) of the main text for all modes. Fig. \ref{fig:rabi_fidelities} displays the Rabi oscillations used to extract the fidelity for mode three.

\subsection{Pulse train calibration}\label{sec:pulse_train_calibration}

\begin{figure*}
    \centering
    \includegraphics[width=0.9\linewidth]{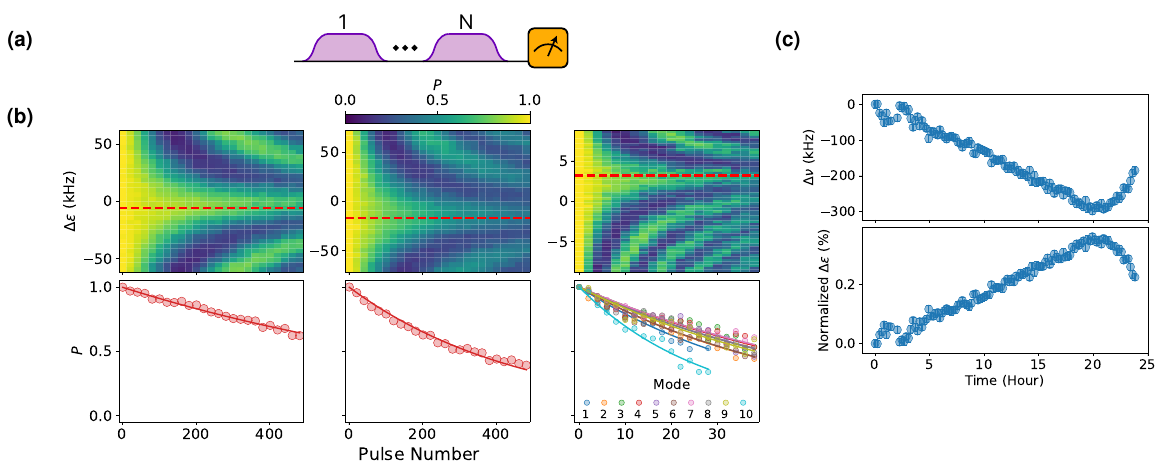}
    \caption{\textbf{Pulse train calibration.} \textbf{(a)} Pulse sequence of the pulse train experiment used to calibrate and characterize the qubit and sideband pulses, where a series of $\pi$-pulses is played. \textbf{(b)} Left column: calibration for $\pi_{ge}$-pulse. 2D plot of the pulse train experiment where we sweep the pulse number and the detuning of the drive. When the pulse is calibrated, there are no oscillations in the pulse train measurement (indicated with a red line). The lower panel displays a pulse train slice at the calibrated drive frequency, showing an exponential decay from which we extract the $\pi$-pulse fidelity to be $\mathcal{F}=\qty[separate-uncertainty = true, multi-part-units=single]{99.931\pm0.002}{\percent}$. Middle column: pulse train calibration for the $\pi_{ef}$-pulse. The extracted fidelity is $\mathcal{F}=\qty[separate-uncertainty = true, multi-part-units=single]{99.769\pm0.004}{\percent}$. Right column: The upper panel displays a pulse train calibration for Mode 3, and the lower panel shows slices of the optimal pulse train calibration for all ten storage cavity modes. \textbf{(c)} To track the drift of the sideband resonance, we perform spectroscopy over 25 hours for the third mode and find that it drifts by up to $\qty{300}{\kilo\hertz}$, corresponding to a drift in the drive amplitude by $\qty{0.4}{\percent}$.}
    \label{fig:pulse_train}
\end{figure*}

Optimization of the sideband pulse involves tuning the pulse duration, amplitude ($\epsilon_{\text{max}}$), and drive frequency. This is performed using a pulse train to amplify rotation errors, where a sequence of $\pi$-pulses is applied, and the $\ket{f}$ state population is measured as a function of the sequence length. We select a target $\pi$-pulse duration that is an integer multiple of the envelope fabric of the Xilinx DAC and perform a fine sweep of the drive amplitude $\epsilon_{\text{max}}$ near the expected $\pi$-pulse amplitude while applying sequences of even (or odd) numbers of $\pi$-pulses, as shown in Fig.~\ref{fig:pulse_train}(b). Given the 16-bit amplitude resolution, this enables a fine sweep of the pulse area. Similar control can also be achieved by adjusting the ramp time; however, this also affects the adiabaticity condition for the pulse.

Since the pulse amplitude also influences the optimal detuning due to the Stark shift, determining the true optimal pulse requires tuning both the amplitude and detuning. Our current experiments were affected by $\sim0.1\%$ per hour drifts in the pulse amplitude, which we attribute to ambient temperature fluctuations impacting the gain of the 5 Watt amplifier used on the sideband drive line. The resulting fluctuations in the sideband resonance frequency over time are shown in Fig.~\ref{fig:pulse_train}(c). These drifts made the two-dimensional sweep of both $\epsilon_{max}$ and $\omega_{sb}$ nonviable.  To mitigate this, we instead relied on a one-dimensional sweep of the pulse amplitude near the optimal detuning, which makes the pulse error quadratic in the detuning error. The optimal calibration is determined by selecting $\epsilon_{\text{max}}$ that minimizes the residual error to an exponential fit.

A comparison of the fidelities extracted from the pulse train calibration and those inferred from Rabi decay are shown in Fig. 2(f) of the main text. The pulse train fidelities are slightly below those inferred from Rabi decay because the pulse train captures leakage effects associated with the pulse profile, such as the ramp. As a result, the Rabi decay fidelities measure the interaction fidelity while the pulse train fidelities measure the pulse fidelity and includes leakage errors. Our pulse profiles can be further optimized to mitigate this.

\begin{figure}
    \centering
    \includegraphics[width=\linewidth]{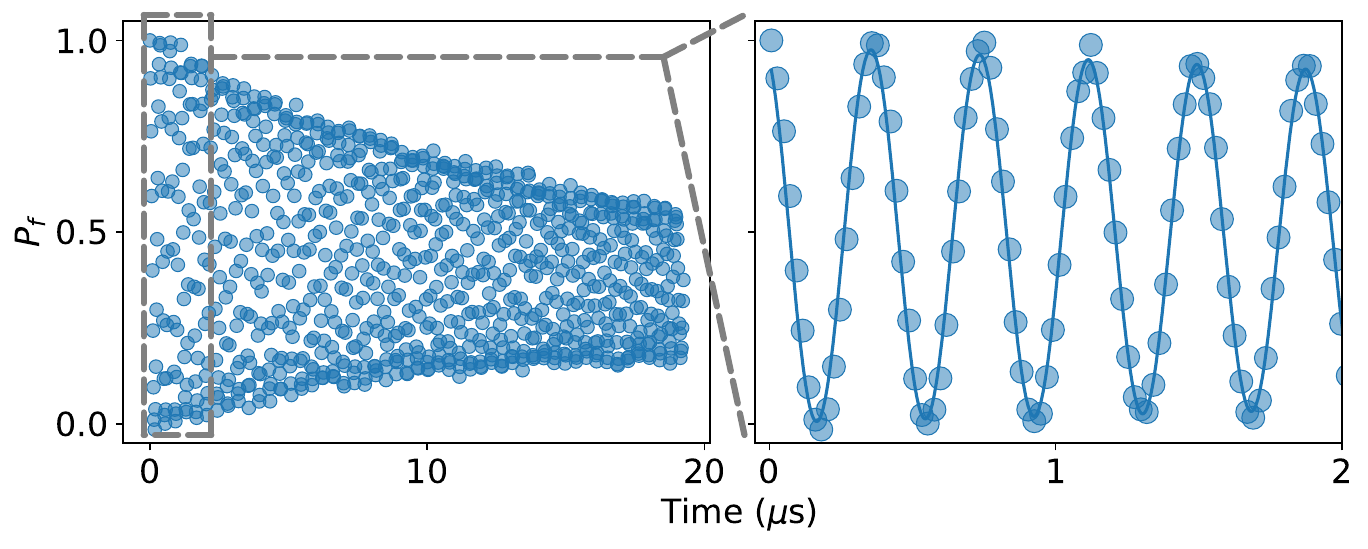}    \caption{\textbf{Sideband interaction fidelity.} The sideband Rabi oscillations (third mode shown) are fit to Eq. \ref{eq:rabi_fidelity} to extract the decay and dephasing times. The evolution for the first $\qty{2}{\micro\second}$ are plotted on the right.}
    \label{fig:rabi_fidelities}
\end{figure}

\section{Floquet analysis of sideband interactions}\label{sec: floquet analysis}

In this section, we explain the numerical procedure used to simulate the sideband interaction. In our simulations, we model the drive signal as a classical monochromatic drive on the transmon, as in Eq. \ref{eq:classical_drive}. To compute the resonance frequency and rate, we first compute the Floquet quasienergy spectrum. The sideband interaction resonance manifests as an avoided crossing in the spectrum, which we fit to extract the frequency and rate. This can be understood in analogy with a driven two-level system, which we detail below. Finally, we use the Floquet formalism to understand the role of the ramp in reducing leakage out of the interaction subspace. This is especially relevant for strongly-driven transmon interactions that lead to significant hybridization between the transmon levels. We derive bounds on the ramp timescale that are required to achieve an optimal fidelity interaction. 

\subsection{Overview of Floquet theory} 

The Floquet formalism provides solutions for the time dynamics of systems whose Hamiltonian $\mathcal{H}(t)$ is periodic with period $T$, i.e., $\mathcal{H}(t) = \mathcal{H}(t + T)$. It is particularly useful for analyzing periodically driven systems, as it captures the effects of strong drives beyond the perturbative regime and the rotating wave approximation (RWA).

The unitary time-evolution operator is given by $U(t_2, t_1)=\mathcal{T}\exp\left(-i\int_{t_1}^{t_2}\mathcal{H}(t)dt\right)$, where $\mathcal{T}$ is the time-ordering operator. When $\mathcal{H}(t)$ is periodic, it can be written in the form
\begin{equation}\label{eq:floquet_unitary_decomp}
    U(t_2,t_1) = e^{-iK[t_0](t_2)}e^{-i\mathcal{H}_F[t_0]\times(t_2-t_1)}e^{iK[t_0](t_1)}.
\end{equation}
Here, we have decomposed the dynamics within a rotating frame given by $U(t)=e^{iK[t_0](t)}$, where $K[t_0](t)$ is the Floquet kick operator. In this frame, the dynamics are generated by a time-independent Floquet Hamiltonian $\mathcal{H}_F[t_0]$. The Floquet Hamiltonian and the kick operator are defined as
\begin{gather}
    e^{-i\mathcal{H}_F[t_0]T} = U(t_0+T,t_0) \\
    e^{-iK[t_0](t)} = U(t, t_0)e^{i\mathcal{H}_F[t_0]\times(t-t_0)},
\end{gather}
where $U(t_0+T,t_0)$ is the unitary over one period. The brackets are used to indicate that the terms are gauge-dependent on the reference stroboscopic time $t_0$. The kick operator has the same periodicity as the Hamiltonian ($K[t_0](t+T) = K[t_0](t)$) and $e^{-iK[t_0](t)}=I$ at stroboscopic times $t=t_0+nT$, where $n$ is an integer. The Floquet modes are defined as the eigenstates of the unitary over one period:
\begin{equation}
    U(t_0+T, t_0)\ket{\Phi_m[t_0]} = e^{-i\varepsilon_mT}\ket{\Phi_m[t_0]}.
\end{equation}
Here, $\ket{\Phi_m[t_0]}$ is the Floquet mode indexed by $m$ and $\varepsilon_m$ is its corresponding quasienergy. We compute the Floquet modes and quasienergies by numerically diagonalizing the unitary over one period. 
 \begin{figure*}
    \centering
    \includegraphics[width=\linewidth]{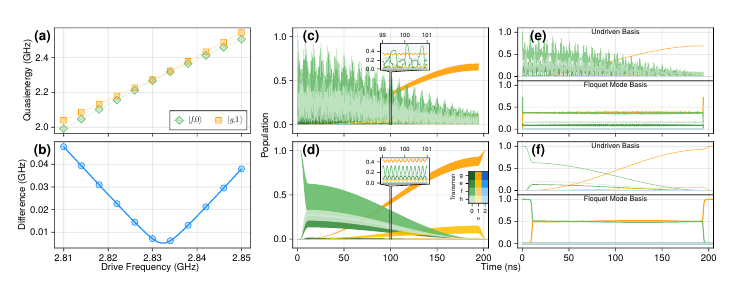}
    \caption{\textbf{Numerical characterization of the sideband interaction.} \textbf{(a)} Quasienergies of the Floquet modes $\ket{f,0}_\text{F}$ and $\ket{g,1}_\text{F}$ for mode three as a function of drive frequency. \textbf{(b)} The fit of the difference in quasienergies is used to extract the resonance frequency and interaction rate. \textbf{(c)} Population evolution in the undriven basis for an $\ket{f0}$-$\ket{g1}$ sideband with a ramp time of $\qty{0.5}{ns}$, showing significant leakage. The inset shows the micro-oscillations in the populations. \textbf{(d)} Corresponding evolution for an $\ket{f0}$-$\ket{g1}$ sideband with a ramp time of $\qty{11.6}{ns}$. \textbf{(e)} Projection of the populations at stroboscopic times into the undriven Hamiltonian basis (top) and the Floquet mode basis (bottom) for a $\qty{0.5}{ns}$ ramp time. With such a short ramp, the initial state starts in a superposition of many Floquet modes, including those outside of the interaction subspace. \textbf{(f)} Corresponding evolution for a ramp time of $\qty{11.6}{ns}$. Here, the ramp initializes the state into an equal superposition of $\ket{f,0}_\text{F}$ and $\ket{g,1}_\text{F}$ (see Section \ref{sec:floquet_ramp}), minimizing leakage and off-resonant rotation.}
    \label{fig:FloquetPlot}
\end{figure*}

Sideband interactions are analogous to Rabi oscillations within a two-state subspace. In this subspace, the Hamiltonian is described by a driven two-level system:
\begin{equation}\label{driven_qubit}
    \mathcal{H} = \frac{\omega_q}{2}\sigma_z + \Omega(e^{-i\omega_dt}\sigma_+ + e^{i\omega_dt}\sigma_-).
\end{equation}
The Floquet Hamiltonian can be derived by transforming into the rotating frame given by $U(t)=e^{-i\frac{\omega_d}{2} (t-t_0)(\sigma_z-I)}$. This yields
\begin{equation}
\mathcal{H}_F[t_0] = \frac{\Delta}{2} \sigma_z + \frac{\omega_d}{2} + \Omega \left(e^{i\omega_d t_0} \sigma+ + e^{-i\omega_d t_0} \sigma_- \right), \end{equation}
where $\Delta = \omega_q - \omega_d$. The eigenfrequencies of this Hamiltonian now correspond to the Floquet quasienergies:
\begin{equation} 
\varepsilon_{\pm} = \frac{1}{2} \left(\omega_d \pm \sqrt{4\Omega^2 + (\omega_q - \omega_d)^2} \right) \quad \text{mod } \omega_d. 
\end{equation}
The strength and resonance frequency of the sideband interaction are determined by identifying the two relevant interacting Floquet modes in the quasienergy spectrum. As in the two-level case, the size of the avoided crossing ($2\Omega$) gives the interaction strength, while the resonance frequency corresponds to the drive frequency at which the gap is minimized (Fig.~\ref{fig:FloquetPlot}(a) and (b)).

\subsection{Unitary evolution of drives with smooth ramping}\label{sec:floquet_ramp}

Smoothly ramping up (and down) the drive strength mitigates leakage out of the interaction subspace when compared to abruptly turning on the drive, such as when using a square pulse. Eq. \ref{eq:floquet_unitary_decomp} provides a way to understand the first of two roles played by the ramp. The first, when its timescale is much longer than the Hamiltonian period ($\tau_{\text{ramp}}\gg T$), is to adiabatically evolve the undriven eigenstates of the system into their corresponding Floquet modes (and back, on the ramp down). By noting that the Floquet modes are eigenstates of $\mathcal{H}_F[t_0]$, this adiabatic transitioning of the undriven interaction subspace to its corresponding subspace in the Floquet mode basis prevents leakage. 

The second role of the ramp is to initialize the starting state into an equal superposition of Floquet modes within the interaction subspace, which ensures that the Rabi oscillations are full contrast. This is performed by diabatically traversing a Landau-Zener transition among the Floquet modes. 

For the sideband interaction, the interaction subspace is $\{\ket{f,n}_\text{F}, \ket{g,n+1}_\text{F}\}$, where the subscript is used to indicate the corresponding Floquet mode. Here, the Landau-Zener velocity is $v_\text{LZ}\sim|\omega_\text{SS}|/(2\pi\tau_\text{ramp})$, where $\omega_\text{SS}=\omega_\text{sb}(\epsilon=\epsilon_\text{max})-\omega_\text{sb}(\epsilon=0)$ is the Stark shift of the sideband resonance at maximum drive amplitude. The avoided crossing energy is $2g_\text{sb}$, where $g_\text{sb}$ is the sideband interaction rate. Hence, the Landau-Zener diabaticity can be defined as 
\begin{equation}
    \Gamma_\text{LZ}=\frac{2\pi|\omega_\text{SS}|}{\tau_\text{ramp}g_\text{sb}^2}=\frac{4\pi\Delta^2}{\tau_\text{ramp}|K|g^2},
\end{equation}
where we have used Eq. \ref{eq:g_sb_f0g1} for $g_{\text{sb}}$ and $\omega_{\text{SS}}\approx -2K\xi^2$, derived from perturbation theory. Since the avoided crossing must be traversed diabatically, $\Gamma_\text{LZ}$ must be large. Hence, the time of the ramp $\tau_\text{ramp}$ must obey the conditions
\begin{equation}\label{eq:ramp_bounds}
   \frac{2\pi}{\omega_\text{sb}} \ll \tau_\text{ramp} \ll \frac{4\pi\Delta^2}{|K|g^2},
\end{equation}
where the lower bound prevents leakage out of the interaction subspace and the upper bound ensures full contrast oscillations. The upper bound can be eliminated by chirping the frequency of the drive so that it always remains on resonance with the interaction. These two roles played by the ramp are illustrated in Fig. \ref{fig:FloquetPlot}. For a ramp time that is too short (Fig. \ref{fig:FloquetPlot}(e)), the initialization is not sufficiently adiabatic to evolve the undriven eigenstates to their corresponding Floquet modes. This is seen in the initialized state being a superposition many Floquet modes, including those outside of the interaction subspace. Fig. \ref{fig:FloquetPlot}(f) shows the evolution using a ramp that satisfies Eq. \ref{eq:ramp_bounds}. Here, the state is initialized into an equal superposition of $\ket{f0}_\text{F}$ and $\ket{g1}_\text{F}$ which leads to an optimal fidelity interaction.

\section{Wigner tomography}\label{sec:wigner_tomography}
\begin{figure}
    \centering
    \includegraphics[width=0.5\textwidth]{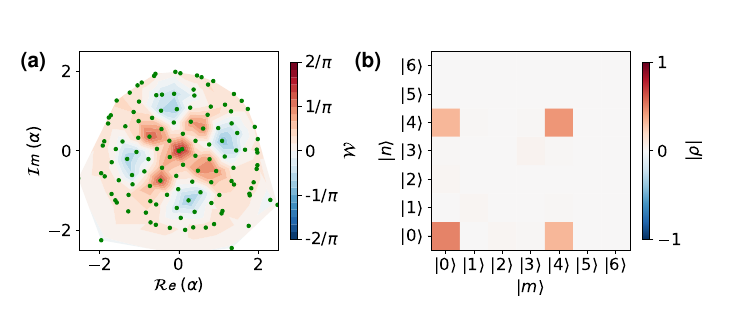} 
    \caption{\textbf{Wigner tomography.} \textbf{(a)} Green points are Wigner displacements chosen to minimize the condition number for a chosen Hilbert space cutoff, shown here along with the Wigner data for $\frac{\ket{0} + \ket{4}}{\sqrt{2}}$. \textbf{(b)} The absolute value of the reconstructed density matrix after applying an inversion protocol that enforces positivity and preserves unit trace. }
    \label{SFig:Wignerpts and Reconstruction}
\end{figure}

We perform single-mode Wigner tomography with slight modifications of the standard protocol consisting of a cavity displacement before a parity measurement ($\hat{\Pi}$). The parity measurement is performed using a transmon $\ket{g}$-$\ket{e}$ Ramsey sequence after idling for a time $1/(2\chi_i)$, where $\chi_i$ is the dispersive shift of the mode. This results in the phase of the transmon superposition state precessing by odd and even multiples of $\pi$ for the corresponding photon number parity, which is mapped to the transmon state by a final $\pi/2$ pulse. 

 Ideally, the overall sequence corresponds to the measurement of the displaced Wigner operator, $\hat{\mathcal{W}}(\alpha) = \hat{\mathcal{D}}_{\alpha}\hat{\Pi}\hat{\mathcal{D}}_{-\alpha}$. The cavity displacements are chosen at an optimal set of points ($\alpha_i$) and the corresponding Wigner operator measurements are given by $x_i = \langle \langle \mathcal{W}(\alpha_i)| \rho \rangle \rangle$. We construct a matrix $\mathcal{M}$ with $\mathcal{M}_{ij} = \left\langle\langle\mathcal{W}(\alpha_i)\right|_j$ that represents measurements of the Wigner operator at all the displacements, which is inverted while imposing positivity and unit trace constraints, thereby extracting the most likely physical density matrix corresponding to the cavity state~\cite{reinhold2019controlling}.  The number of columns of $\mathcal{M}$ is $d^2$, where $d$ is the truncated dimension of the Hilbert space of the cavity up to which the tomography is valid, while the rows correspond to the number of Wigner measurements ($>d^2$).  The displacements are chosen to minimize the error from the inversion, corresponding to minimizing the ratio of the largest to the smallest eigenvalue of $\mathcal{M}$ (the condition number). 

\subsection{Error mitigation of parity measurement}

\begin{figure}
    \centering
    \includegraphics[width=\linewidth]{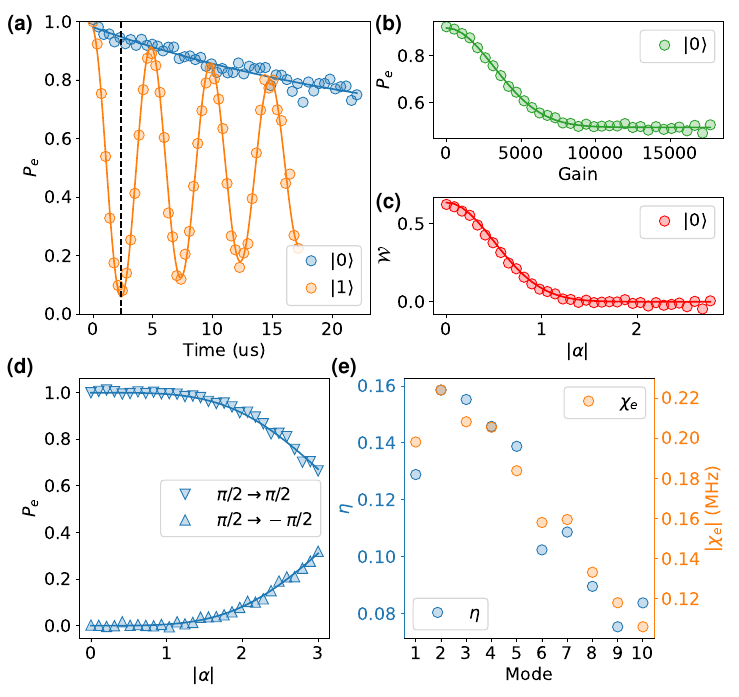}
    
    \caption{\textbf{Parity measurement error mitigation and contrast correction.} \textbf{(a)} Transmon Ramsey experiment with $\frac{\pi}{2}$ pulses with identical phases following initializing the cavity in the vacuum or $\ket{1}$ state. The dotted black line shows the $1/(2\chi)$ time. \textbf{(b)} Transmon $\ket{e}$ probability ($P_e$) after the Wigner tomography sequence, as a function of the gain of the gaussian cavity displacement pulse, fit to the functional form $P_e (a) = (1 + e^{-2(sa)^{2}})(p_e - p_g)/2+ p_g$.  \textbf{(c)} Wigner function versus $\alpha$ for $\ket{0}$ of mode 3 after rescaling the measurement using Eqn.~\ref{Wigner0_calibration}. \textbf{(d)} Contrast reduction of a parity measurement as a function of $\alpha$ for mode 3 arising from dispersive frequency shift of the transmon and finite pulse bandwidth, fit to the functional form described in the text. \textbf{(e)} The fit value of $\eta$ is $\propto\chi_e$ as expected.}
    \label{SFig:Parity Calibration}
\end{figure}

For weak coupling, the Ramsey time for the parity measurement can constitute a significant fraction of the transmon decoherence time, leading to errors in the parity measurement caused by the transmon's decay and dephasing. This can be seen in Fig. \ref{SFig:Parity Calibration}(a) where we show a Ramsey experiment after preparing the cavity in $\ket{0}$ or $\ket{1}$. However, since this time is much shorter than the cavity decay time, the resulting error can be calibrated and corrected. This approach represents a form of ancilla error mitigation that enables the recovery of parity measurement expectation values and statistics.

In the following, we rigorously demonstrate the feasibility of this error mitigation strategy by modeling the decoherence during the parity measurement using a Lindblad master equation
\begin{equation}
    \odv{\rho(t)}{t} = -i[\mathcal{H}, \rho(t)] + \sum_{i=0}^{1} \left( L_i\rho(t)L_i^\dag - \frac{1}{2} \{L_i^\dag L_i, \rho(t)\}\right),
\end{equation}
where 
\begin{gather}
    \mathcal{H} = \frac{\chi}{2}b^\dag b \sigma_z \\ L_0 = \sqrt{\gamma}\sigma_- \\
    L_1 = \sqrt{\frac{\gamma_\varphi}{2}} \sigma_z.
\end{gather}
Here, $\gamma$ is the decay rate and $\gamma_\varphi$ is the pure dephasing rate of the qubit. Our system evolves under the dispersive Hamiltonian. The general time evolution of the density matrix can be solved to give
\begin{align}
\begin{split}
    \rho(t) = \sum_{n_1 n_2}\biggl(&\rho^{ee}_{n_1n_2}(t) \ket{en_1}\bra{en_2}  \\
    &+ \rho^{eg}_{n_1n_2}(t) \ket{en_1}\bra{gn_2} \\
    &+ \rho^{ge}_{n_1n_2}(t) \ket{gn_1}\bra{en_2} \\
    &+ \rho^{gg}_{n_1n_2}(t)\ket{gn_1}\bra{gn_2}\biggr)
\end{split}
\end{align}
where 
\begin{gather}
    \rho^{ee}_{n_1n_2}(t) = e^{-\gamma t}e^{-i\frac{\chi}{2}(n_1-n_2)t}\rho^{ee}_{n_1n_2}(0) \\
    \rho^{eg}_{n_1n_2}(t) = e^{-(\frac{\gamma}{2} + \gamma_\varphi) t}e^{-i\frac{\chi}{2}(n_1+n_2)t}\rho^{eg}_{n_1n_2}(0) \\
    \rho^{ge}_{n_1n_2}(t) = e^{-(\frac{\gamma}{2} + \gamma_\varphi) t}e^{i\frac{\chi}{2}(n_1+n_2)t}\rho^{ge}_{n_1n_2}(0) \\
    \begin{split}
        \rho^{gg}_{n_1n_2}(t) &= \left( \rho^{gg}_{n_1n_2}(0) + \frac{\gamma \rho^{ee}_{n_1n_2}(0)}{\gamma + i\chi(n_1-n_2)}\right)e^{i\frac{\chi}{2}(n_1-n_2)t} \\
        &\phantom{=} \ - \frac{\gamma \rho^{ee}_{n_1n_2}(0)}{\gamma + i\chi(n_1-n_2)}e^{-\gamma t}e^{-i\frac{\chi}{2}(n_1-n_2)t}
    \end{split}
\end{gather} 
and we have used $t=0$ as our reference time. Assuming the qubit is in $\ket{g}$ and disentangled with the cavity, we can perform a parity measurement by applying a $\frac{\pi}{2}$-pulse, waiting $\tau=\frac{1}{2\chi}$, and then applying another $\frac{\pi}{2}$-pulse. Here, we can see that the probability of measuring the qubit in $\ket{e}$ is
\begin{equation}
    P_e = \frac{1}{2}\left(1 - e^{-(\frac{\gamma}{2} + \gamma_\varphi)t}\right) + e^{-(\frac{\gamma}{2} + \gamma_\varphi)t}P_{\text{even}}.
\end{equation}

\begin{figure*}[t]
    \centering
    \includegraphics[width=\linewidth]{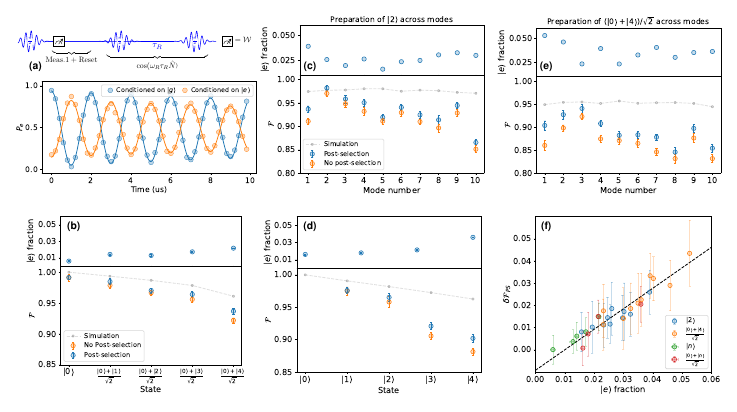}
    \caption{\textbf{Wigner tomography with post-selection.}  \textbf{(a)} Validation of post-selection (without cavity displacement), showing the outcome of the second transmon measurement conditioned on the first measurement result while sweeping the idle time $\tau_R$ in a subsequent Ramsey experiment. The transmon is initially prepared in the state $\frac{\ket{g}+\ket{e}}{\sqrt{2}}$ (pulse sequence shown at the top). \textbf{(b)} Wigner tomography results with and without post-selection for the preparation of the state $\frac{\ket{0}+\ket{n}}{\sqrt{2}}$ in mode 3. (Top) The fraction of $\ket{e}$ outcomes in the first measurement as a function of $n$. (Bottom) Comparison of state fidelity from Wigner tomography as a function of $n$, both without post-selection and with post-selection on the transmon being in $\ket{g}$ at the end of the first measurement. Corresponding results for (top) $\ket{e}$ outcomes in the first measurement and (bottom) state fidelities with and without post-selection (on $\ket{g}$) for \textbf{(c)} the $\ket{n}$ preparation sequence applied to mode 3, \textbf{(d)} the preparation of $\ket{2}$, and \textbf{(f)} the preparation of $\frac{\ket{0}+\ket{4}}{\sqrt{2}}$ across the first ten cavity modes. \textbf{(f)} Improvement in state-fidelity with and without post-selection $\delta\mathcal{F}_{PS}$ as a function of the fraction of shots in $\ket{e}$ in the first post-selection measurement across all the measurements. }
    \label{fig:wigner_tom_post_selection}
\end{figure*}

Hence, under decay and dephasing of the qubit we can still recover the parity measurement statistics by offsetting and rescaling the qubit measurement outcomes. Consistent with the expression derived above, we convert between the transmon measurement $P$ and the expected parity measurement $\hat{\Pi}$ using $\hat{\Pi} = \frac{P-p_g}{p_e - p_g}$.

The $p_g, p_e$ values may be calibrated by measuring the decay in contrast of a transmon Ramsey sequence. For our device, due to the spurious dispersively coupled mode excited during transmon readout (see Section \ref{sec:spurious_mode}), the Ramsey experiment exhibits small micro-oscillations that lead to an additional few percent error if we rely solely on the exponential fit. We therefore estimate \( p_g \) and \( p_e \) by using the Wigner function measurement for the vacuum state of each mode as a calibration. The displacement pulses used in the tomography protocol are $4\sigma$ Gaussian pulses with $\sigma = 50~\mathrm{ns}$ for most modes (125 ns for modes 8, 9, and 10). We achieve different cavity displacements ($\alpha$) by adjusting the pulse gain and phase. We fit the measured transmon probability versus gain ($P(a)$) to extract the scaling ($\beta = \alpha/a$) between the displacement and the pulse gain, as well as $p_g$ and $p_e$, as shown in SFig. 6(b). We subsequently convert between $P(a)$ and $\mathcal{W}(\alpha)$ using:
\begin{equation}
    \mathcal{W}(\alpha) = \mathcal{W}(\beta a) = \frac{2}{\pi}\left(2\left(\frac{P(a) - p_g}{p_e - p_g}\right) - 1\right)
    \label{Wigner0_calibration}
\end{equation}

In addition to decoherence, the parity measurement also has a systematic error arising from the finite bandwidth of the parity-measurement $\pi/2$ pulse~\cite{chakram2020multimode}. This leads to a reduction in the contrast in the measurement of the Wigner function for larger cavity displacements, which we calibrate using a sequence comprising two back-to-back $\pi/2$ pulses with either the same or opposite phase, without waiting time between. The result of such a measurement for a given mode (2) is shown in SFig.6(c). The reduction in contrast is consistent with that expected from the off-resonant error in the $\pi/2$ pulse from a frequency shift of $\chi\alpha^2$. The curves are fit to a functional form $C(\alpha) = \frac{1}{1+\eta^2 \alpha^4}$ and $1-C(\alpha)$. Here $\eta = \chi/(2\epsilon)$ where $\epsilon$ is the strength of the qubit Rabi drive. The $\eta$ extracted from the fit to the parity measurement is found to be proportional to $\chi$, as shown in SFig. 6(d). 

We ignore the additional readout error arising from the cross-Kerr interaction between the storage and readout modes~\cite{chakram2020multimode}, since this is sufficiently small.   We compensate for this effect by scaling the measurement of the Wigner operator for a given state and displacement using a linear transformation ($\mathcal{W}(\alpha,\rho) \rightarrow \mathcal{W}(\alpha,\rho)/C(\alpha)$). We note that this calibration was empirically found to only change the fidelity by a few percent.

The reconstruction procedure described above provides the best estimate of the cavity density matrix while mitigating errors in the parity measurement caused by transmon decay and coherent pulse errors. However, several sources of error can still affect the reconstructed $\rho$, including random measurement noise, uncertainty in the calibration of the parity measurement ($p_e, p_g$), and errors arising when the transmon is not in the $\ket{g}$ state at the start of the measurement. For the first two sources of error, we estimate the resulting uncertainty in the fidelity and add it to our error bar. To mitigate the latter error, we perform a transmon measurement prior to Wigner tomography and post-select based on the result. We detail this error analysis in the following sections.

\begin{figure*}[t]
    \centering
    \includegraphics[width=0.9\linewidth]{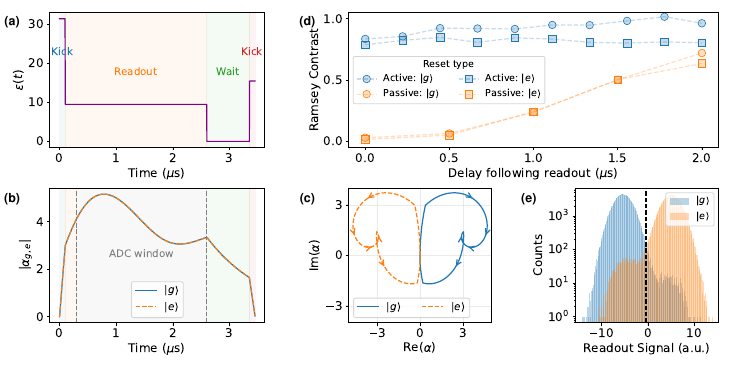}
    \caption{\textbf{Integrated readout and reset pulse.} \textbf{(a)} Pulse sequence for integrated readout and readout reset pulse. \textbf{(b)} Evolution of the readout mode's displacement magnitude for each of the transmon states. \textbf{(c)} Evolution of the readout mode in phase space for each transmon state. \textbf{(d)} Characterization of the readout mode photon population using a transmon Ramsey measurement, illustrating that the reset protocol clears readout photons much faster than passive reset. \textbf{(e)} Histogram of readout signal along the rotated quadrature axis that most distinguishes the single-shot readout histograms, with the threshold line corresponding to a readout fidelity of $97.5\%$.}
    \label{fig:readout_reset}
\end{figure*}

\subsection{Wigner tomography with post-selection}

The standard protocol for Wigner tomography relies on the transmon being in the $\ket{g}$ state and initially disentangled from the cavity state. Although the sideband-based state preparation sequences we implement are ideally designed to achieve this, transmon and cavity decoherence and coherent calibration errors can result in the transmon and the cavity modes remaining weakly entangled at the end of state preparation. To mitigate the resulting uncontrolled tomography error, we implement a protocol for Wigner tomography with post-selection. Here, we perform an initial transmon measurement before Wigner tomography and post-select outcomes, where the first measurement confirms that the transmon is in the $\ket{g}$ state. The pulse sequence corresponding to this protocol is shown in Fig.~\ref{fig:wigner_tom_post_selection}(a).

We test the efficacy of post-selection in the context of Wigner tomography through an experiment consisting of a $\pi/2$ pulse, an initial post-selection measurement and readout cavity reset, followed by a Ramsey sequence and a second transmon measurement. The results of this experiment, conditioned on the first measurement outcome being $\ket{g}$ or $\ket{e}$, are shown in Fig.~\ref{fig:wigner_tom_post_selection}(b). We observe the expected anti-correlated oscillations, and a drop in contrast for an $\ket{e}$ outcome in the first measurement from transmon decay during the measurement and readout reset.

We present a comparison of the measured fidelities with and without post-selection for the preparation of $\ket{n}$ and $\frac{\ket{0}+\ket{n}}{\sqrt{2}}$ superposition states for a given mode (3) in Fig.~\ref{fig:wigner_tom_post_selection}(c) and (d).  We also show the corresponding data for state preparation of Binomial code states across the 10 modes of our system in Fig.~\ref{fig:wigner_tom_post_selection}(e) and (f).  We see that across all these state preparation experiments,  there is an increase in the Wigner tomography fidelity with post-selection linearly correlated with the fraction of shots measured in $\ket{e}$, as shown in Fig.~\ref{fig:wigner_tom_post_selection} (g). 

\subsection{Fast reset of the readout cavity}
 To minimize infidelity from cavity decoherence during post-selection, we developed a fast, integrated readout and reset pulse similar to the CLEAR pulse \cite{mcclure2016rapid}. This pulse is particularly necessary due to our relatively low readout $\kappa\approx 1/(600~\mathrm{ns})$. The schematic of the pulse is shown in Fig. \ref{fig:readout_reset}(a), with a carrier frequency positioned between the readout resonator frequencies corresponding to the $\ket{g}$ and $\ket{e}$ states. The evolution of the magnitude of the cavity displacement and the cavity state in phase space are illustrated in Fig. \ref{fig:readout_reset}(b) and (c), respectively.

The pulse consists of an initial ring-up segment that rapidly drives the cavity to the target displacement, followed by a readout segment during which the cavity settles to the nominal values for each state ($\alpha_{g,e} \approx \frac{2\epsilon}{\kappa \pm i\chi}$), corresponding to angles $\theta_{e/g} = \pm\arctan{\frac{\chi}{\kappa}} = \pm\theta_r$ about the imaginary axis. To reset the cavity states corresponding to both transmon states, the reset pulse includes a wait period during which the cavity states rotate in opposite directions, converging on the opposite side of the cavity phase space. This is followed by a final pulse that pushes both states back to the origin. The total reset time is given by $\tau_r = \frac{2(\pi-\arctan{\frac{\chi}{\kappa}})}{\chi} \approx 0.7~\mu$s.

We optimize the amplitude and duration of each component of the readout and reset sequence using the contrast of a transmon Ramsey experiment performed immediately after the readout and reset pulse, along with readout histogram measurements. This optimization minimizes transmon dephasing caused by residual readout photons while simultaneously maximizing readout fidelity. The contrast following the optimized readout pulse is compared with that of a regular uniform readout pulse in Fig. \ref{fig:readout_reset}(d). We observe that the contrast remains near its full value immediately after readout for the active reset pulse, unlike the uniform (passive) readout pulse. Additionally, we include an extra $\qty{2}{\micro\second}$ after the readout and active reset pulse to allow any residual readout photons to fully leak out of the cavity.

With this readout pulse, we are able to achieve a readout fidelity of $97.5\%$ without the use of a parametric amplifier, with an integration time of approximately $1.5~\mu$s, the corresponding histogram for which is shown Fig. \ref{fig:readout_reset}(e).  This corresponds to a $1.5\%\text{ } (0.5\%)$ assignment error for the transmon measured in $\ket{e} (\ket{g})$. We note that we may further reduce the assignment error to $<0.5\%$ by post-selecting shots that lie in regions were the assignment error is below that threshold. This necessitates throwing away nearly $50\%$ of the measurements where the first measurement is $\ket{g}$, coming at the expense of increased reconstruction errors from increased measurement noise. This compromise, in conjunction with $> 95\%$ of shots being in $\ket{g}$ resulted in no improvement in Wigner tomography fidelity with further reducing assignment error. We therefore chose to perform post-selected Wigner tomography experiments with the standard threshold voltage, indicated by the dashed black line in Fig \ref{fig:readout_reset}(e).

\begin{figure*}[t]
    \centering
    \includegraphics[width=0.97\linewidth]{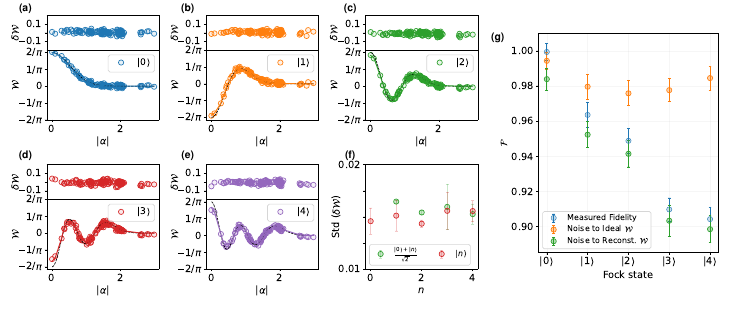}
    \caption{\textbf{Wigner tomography error analysis.} \textbf{(a-e)} Comparison of measured Wigner functions (circle markers) with those of the reconstructed $\rho$ (solid colored lines) and ideal Wigner functions (dashed black lines) as a function of $|\alpha|$ for Fock states ($n = 0$ to $4$) in mode 3 of the multimode register. \textbf{(f)} Standard deviation of the difference $\delta\mathcal{W}$ between the measured and reconstructed Wigner functions for data corresponding to the preparation of $\ket{n}$ and $\frac{\ket{0} + \ket{n}}{\sqrt{2}}$ in mode 3. \textbf{(g)} Fidelity and spread obtained from reconstruction following the addition of random noise to ideal and reconstructed Wigner functions.}
    \label{wig_tom_error}
\end{figure*}

\subsection{Wigner tomography error analysis}\label{wigner error}

In addition to errors from transmon excitation, Wigner tomography infidelity also arises from measurement noise and fluctuations in the calibration of the parity measurement ($p_g, p_e$ values). We estimate measurement noise by comparing the measured Wigner function with the expected Wigner function for the reconstructed $\rho$ at corresponding phase-space points. Their comparison and their difference for Fock states up to $n = 4$ in mode 3 are shown in Fig. \ref{wig_tom_error}(a–e). The standard deviation of the noise, plotted as a function of $n$ for Wigner functions of both $\ket{n}$ and $\frac{\ket{0} + \ket{n}}{\sqrt{2}}$ in Fig. \ref{wig_tom_error}(f), shows that the error remains largely consistent across states.

To quantify the fidelity error due to measurement noise, we introduce random noise with the same standard deviation at each data point and perform the reconstruction multiple times (50 iterations). This noise addition is applied to both an ideal Wigner function for $\ket{n}$ and the Wigner function corresponding to the reconstructed $\rho$ from our measurements, with the results shown in Fig. \ref{wig_tom_error}(g).  Reconstruction following the addition of noise to the ideal Wigner function leads to a drop in the fidelity, as expected. Measurement noise, similarly leads to an underestimate in the fidelity of our measured state by $\sim 2\%$. While the spread in the fidelity is similar for both cases, we take the spread in the fidelity from noise added to the reconstructed Wigner function as a proxy for the error in the fidelity of our reconstructed state in the results presented here. 

Finally, we account for fidelity error due to fluctuations in the parity measurement calibration, specifically in the values of $p_e$ and $p_g$ used to rescale the parity measurement. These fluctuations arise from variations in the Ramsey contrast occurring over the timescale of a Wigner tomography measurement. To quantify this effect, we determine the spread in $p_e$ and $p_g$ through repeated calibrations of the vacuum state and estimate the corresponding fidelity error. For mode 3, this corresponds to $p_g \pm \delta p_g = 0.075 \pm 0.009$ and $p_e \pm \delta p_e = 0.912 \pm 0.006$ over seven datasets. We observe that $p_g$ exhibits nearly a 1\% variation, which leads to a comparable error in the reconstructed fidelity. To estimate the fidelity error from this effect, we perform the reconstruction for ($p_g \pm \delta p_g$, $p_e \pm \delta p_e$) and evaluate the resulting spread in fidelity. For the data presented in Fig. \ref{fig:wigner_tom_post_selection}, error bars for individual datasets are obtained by adding the errors from measurement noise and parity calibration in quadrature. The final data is presented as a weighted average of all available datasets using the same calibration routine.

\begin{figure*}
    \centering
    \includegraphics[width=\linewidth]{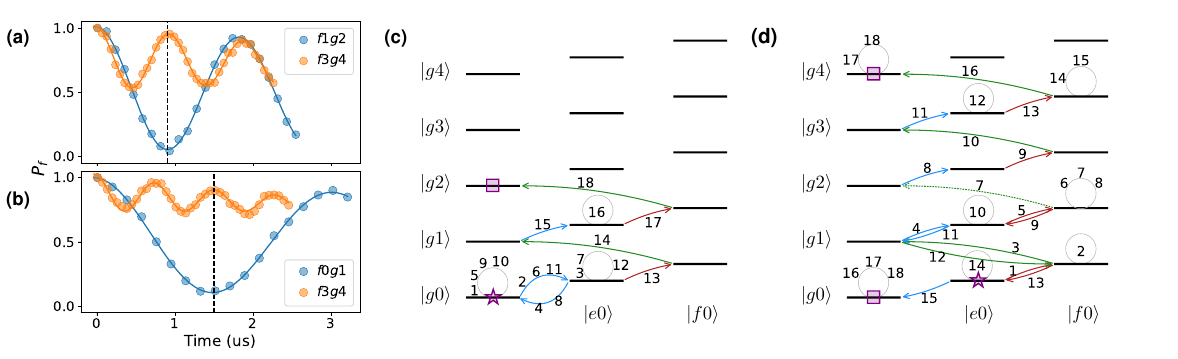}
    \caption{\textbf{Calibration of photon number-selective sideband.} \textbf{(a)} Measured $\ket{f}$ population after applying the sideband on resonance with the $f1-g2$ transition with $g_{\text{sb}}=|\chi_f|/\sqrt{2}$ on states prepared in $\ket{f1}$ (blue) and $\ket{f3}$ (orange). \textbf{(b)} Measured $\ket{f}$ population after applying the sideband drive on resonance with the $\ket{f0}$-$\ket{g1}$ transition with $g_{\text{sb}}=\sqrt{3}|\chi_f|/4$ on states prepared in $\ket{f0}$ (blue) and $\ket{f3}$ (orange). \textbf{(c)} Combined Hilbert space of a transmon and one of the cavity modes, illustrating the state transitions induced by the encoding pulse sequence in Eq. \ref{eq:binomial_enc_unitary}. The purple star indicates the initial state, and the purple square is the final state ($\ket{g}\rightarrow\ket{2}$). The transition numbers correspond to the pulse numbers in Fig. 5(a) of the main text. \textbf{(d)} The state transitions for the encoding operation with the transmon initialized in $\ket{e}$ ($\ket{e}\rightarrow(\ket{0}+\ket{4})/\sqrt{2}$).}
    \label{fig:pnr_sideband}
\end{figure*}

\section{Binomial code encoding gate}

\subsection{Photon-number-selective sideband calibration}\label{sec:pnr_sideband}
The binomial encoding gate maps the transmon states $\{\ket{g}, \ket{e}\}$ to the binomial code states $\{\ket{2}, \frac{\ket{0}+\ket{4}}{\sqrt{2}}\}$ through a sequence of transmon rotations and fast $\ket{f,n}$-$\ket{g,n+1}$ sideband pulses, along with two photon-number-selective sideband pulses.  The latter leverages the photon-number dependence of the $\ket{fn}$-$\ket{g,n+1}$ sideband transition frequency, which is dispersively shifted by $n \chi_f$ as shown in Fig. 2(b) of the main text. The pulse targets a pair of different photon number states and utilizes the drive detuning to adjust the sideband rates to perform a $\pi$-pulse on a given photon number while implementing a $2m\pi$-rotation on the other. This enables photon-number-selective sideband operations, significantly faster than achieved by targeting a single photon-number state, which requires $g_\text{sb} \ll |\chi_f|$ and can be prohibitively slow for weak coupling. The sequence used in this work (shown in Fig. 5 (a) of the main text) is:
\begin{widetext}
\begin{align}\label{eq:binomial_enc_unitary}
\begin{split}
    U_{\text{enc}}&: \pi_{ef} \rightarrow \pi_{ge} \rightarrow \pi_{f0g1} \rightarrow \pi_{ge} \rightarrow \pi_{ef} \rightarrow \pi_{ge} \rightarrow \frac{\pi}{2}_{f1g2} \rightarrow \pi_{ge} \rightarrow \pi_{ef} \rightarrow \pi_{f2g3} \\ 
    &\phantom{{}:{}} \rightarrow \pi_{ge} \rightarrow \pi_{f0g1} \rightarrow \pi_{ef} \rightarrow \begin{cases} \pi_{f0g1} \\ 4\pi_{f3g4} \end{cases}\hspace{-1em} \rightarrow \pi_{ge} \rightarrow \pi_{f3g4} \rightarrow \pi_{ef} \rightarrow \begin{cases} \pi_{f1g2} \\ 2\pi_{f3g4} \end{cases}\hspace{-1em}.
\end{split}
\end{align}
\end{widetext}

The two rate-limiting photon-number-selective pulses implement $\pi_{f0g1}$–$4\pi_{f3g4}$ at a rate $1/\tau = \sqrt{3} |\chi_f|$ and $\pi_{f1g2}$–$2\pi_{f3g4}$ at a rate $1/\tau = 2\sqrt{2}|\chi_f|$. Although these rates depend on the strength of the dispersive interaction, they scale with $\chi_f$ and feature prefactors greater than unity, accelerating selective operations by factors of $3.2$ and $5.2$ compared to $\chi_e$, respectively. This rate can potentially be further enhanced without introducing additional Purcell loss to the cavity by operating the transmon near the straddling regime~\cite{koch2007charge}.

\begin{figure*}
    \centering
    \includegraphics[width=1\linewidth]{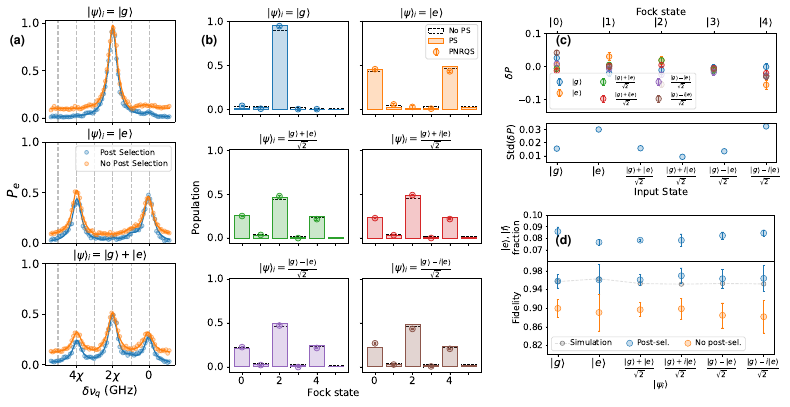}
    \caption{\textbf{Binomial encoding gate characterization.} \textbf{(a)} Photon-number-resolved qubit spectroscopy (PNRQS) with and without post selection after running the encode sequence on $\{\ket{g}, \ket{e}, \frac{\ket{g} + \ket{e}}{\sqrt{2}}\}$, from top to bottom. \textbf{(b)} Comparing the populations extracted from PNRQS and from Wigner tomography (with and without post selection) following acting the encode gate after preparing the transmon in all the cardinal points on the $\{\ket{g}, \ket{e}\}$ Bloch sphere. \textbf{(c)} (Bottom) Improvement in fidelity from Wigner tomography with post selection and (top) the fraction of the shots that are excluded in the post-selection measurement \textbf{(d)} (Top) The differences in the populations extracted from PNRQS and Wigner tomography plotted as a function of the Fock level for the different input states. The standard deviation of the population differences for the different input states are shown in the bottom.}
    \label{fig:BinomialEncodingFigure}
\end{figure*}

Since sideband interactions are analogous to Rabi oscillations in a two-state subspace, the rates above can be derived considering the Hamiltonian of a driven two-level system in Eq. \ref{driven_qubit}. The Rabi oscillation rate for a detuning $\Delta$ from the sideband resonance is given by $\Gamma = \sqrt{g_\text{sb}^2 + \Delta^2/4}$. Consider sidebands acting on two states with differing photon numbers $n_1$ and $n_2$ where $n_2\geq n_1$. Their  resonant sideband rates taken to be $g_\text{sb1}$ and $g_\text{sb2} = \sqrt{\frac{n_2+1}{n_1+1}}g_\text{sb1}$, respectively. By leveraging the faster rate of a detuned Rabi oscillation, we can calibrate $g_\text{sb1}$ to perform a $\pi$-pulse on the first state ($n_1$) and a $2m\pi$-pulse on the second ($n_2$). Driving on resonance with the first state to perform a high-fidelity $\pi$ pulse, we derive that
\begin{equation}\label{photon_num_sel_sb}
    g_\text{sb1} = \frac{(n_2-n_1)|\chi_f|}{2\sqrt{4m^2-(n_2+1)/(n_1+1)}}.
\end{equation}
If we were instead to drive on resonance with the second state (which has a faster sideband rate), we require $g_\text{sb1} = \frac{(n_2-n_1)|\chi_f|}{2\sqrt{4m^2(n_2+1)/(n_1+1)-1}}$, which is strictly slower than driving on the first state. 

We calibrate our photon-number-selective sidebands using Eq. \ref{photon_num_sel_sb} to estimate the required sideband rate. We iteratively perform sideband Rabi oscillations as a function of time while finely sweeping the drive amplitude around the expected value, simultaneously performing spectroscopy to determine the resonance frequency at each amplitude. This process is repeated until the durations of the $\pi$-pulse and $2m\pi$-pulse converge to $\sim1\%$, resulting in a $\sim 0.01\%$ calibration error in the populations.

\begin{figure*}
    \centering
    \includegraphics[width=1\linewidth]{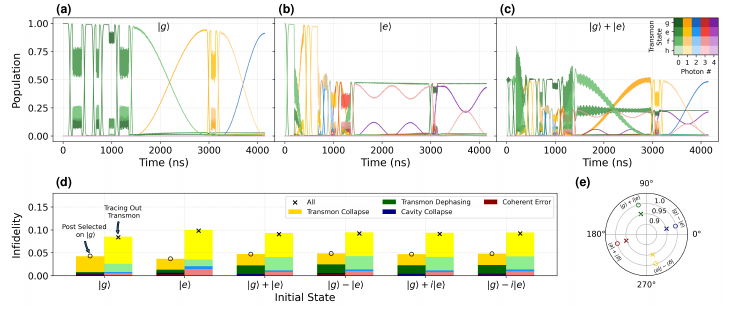}
    \caption{\textbf{Error budget for binomial encoding gate.} \textbf{(a)-(c)} Lindblad master equation simulations for the binomial encoding sequence, starting from the states $\ket{g0}$ (a), $\ket{e0}$ (b), and $\frac{\ket{g0}+\ket{e0}}{\sqrt{2}}$ (c). \textbf{(d)} Error budget for the encoding operation. The primary source of infidelity arises from transmon decay and dephasing, with coherent errors being the next most significant contributor. The contribution from cavity decay is the smallest factor. In the simulations, we ignore cavity decay during the post-selection measurement, which we estimate to contribute an additional $\qty{1}{\percent}$ to the infidelity. We present both the infidelity of the final cavity state post-selected on the transmon being in $\ket{g}$ and after tracing out the transmon. \textbf{(e)} The relative phases between $\ket{0_{L}}$ and $\ket{1_{L}}$ after encoding an initial state of the form $\left(\ket{g}+e^{i\varphi}\ket{e}\right)\ket{0}$ for $\varphi = 0, \pm \pi/2$, and $\pi$. The encoding preserves the relative phases.}
    \label{fig:BinomialEncodingChar}
\end{figure*}

\subsection{Gate characterization and error budget}

The binomial encoding gate is characterized by preparing the transmon in the ${\ket{g},\ket{e}}$ states at six cardinal points on the Bloch sphere and measuring the resulting cavity state after the gate operation. The cavity state is analyzed using Photon Number Resolved Qubit Spectroscopy (PNRQS) and Wigner Tomography, both with and without a transmon post-selection measurement prior to the final measurement. The results of PNRQS with and without post-selection are shown in Fig.~\ref{fig:BinomialEncodingChar}(a). Without post-selection, transmon decay—particularly during the longer photon-number-selective sideband pulses—results in a higher fraction of population remaining in the $\ket{e}$ state. In the PNRQS measurement, this leads to an elevated baseline, which, as expected, is consistent with the excited-state fraction measured in the post-selected Wigner tomography experiment. The PNRQS and Wigner tomography measurements are normalized according to the protocol presented in Section~\ref{Data normalization}. 

From the post-selected PNRQS measurements, we extract the cavity population in each Fock state by fitting the spectrum to a sum of Lorentzians while constraining the linewidths to be identical and the center frequencies to shift with photon number according to an independently measured $\chi$. This corresponds to a measurement of the diagonal elements of the cavity density matrix, which we compare to those extracted from Wigner tomography for each input state in Fig.~\ref{fig:BinomialEncodingChar}(b).

The measured populations from Wigner tomography and PNRQS, post-selected on the transmon being in $\ket{g}$, are found to be consistent, as shown in Fig.~\ref{fig:BinomialEncodingChar}(c), where we plot the differences in the populations extracted from PNRQS and Wigner tomography for each input state. The standard deviation of the populations extracted from the two methods is found to be within $2\%$. This further corroborates the fidelities measured via Wigner tomography, shown in Fig.~\ref{fig:BinomialEncodingChar}(d), both with and without post-selection. As before, the improvement in fidelity with post-selection closely matches the $\ket{e}$ fraction in the post-selection, which is also found to be consistent across Wigner tomography and PNRQS measurements.

\section{NOON state encoding gate}\label{sec:noon_state}
The NOON state encoding gate maps the transmon states $\{\ket{g}, \ket{e}\}$ to the states $\{\ket{N0}, \ket{0N}\}$ in any pair of cavity modes through a sequence of transmon rotations and alternating fast $\ket{f,n}$-$\ket{g,n+1}$ sideband pulses to each mode and is shown in Fig.~4(c) of the main text and reproduced below. The first and second modes are indexed by m1 and m2, respectively: 

\begin{widetext}
\begin{equation}
    U_\text{enc}(N):
        \begin{cases}
        \pi_{ge} \rightarrow \pi_{ef} \rightarrow \pi^{\text{m2}}_{f0\text{-}g1} \rightarrow \pi_{ef} \rightarrow \pi^{\text{m1}}_{f0\text{-}g1} & \text{if $N=1$} \\ 
        & \vspace{0.05in} \\
            \begin{aligned}
            &\pi_{ge} \rightarrow \pi_{ef} \rightarrow \pi^{\text{m2}}_{f0\text{-}g1} \rightarrow \pi_{ef} \rightarrow \pi_{ge} \rightarrow \pi^{\text{m1}}_{f0\text{-}g1} \\
            &\rightarrow \sum_{n=1}^{(N-1)^*} \Bigl( \pi_{ef} \rightarrow \pi_{ge} \rightarrow \pi^{\text{m2}}_{f,n\text{-}g,n+1} \rightarrow \pi_{ef} \rightarrow \pi_{ge}  \rightarrow \pi^{\text{m1}}_{f,n\text{-}g,n+1}\Bigl) \\
            &\rightarrow \pi_{ef} \rightarrow \pi_{ge} \rightarrow \pi^{\text{m2}}_{f,N-1\text{-}g,N}\rightarrow \pi_{ef} \rightarrow \pi^{\text{m1}}_{f,N-1\text{-}gN}
            \end{aligned} & \text{if $N>1$}.
        \end{cases}
\end{equation}
\end{widetext}
Here, the sum indicates the set of pulses, in parentheses, that must be played sequentially by index $n$ and is not inclusive of the maximum index value $N-1$ (indicated by the asterisk). The resulting cavity state is characterized through three methods: photon number resolved qubit spectroscopy (PNRQS), qubit tomography following an encode and decode operation, and multimode Wigner tomography.

\subsection{Multimode Wigner tomography}

\begin{figure*}
    \centering
    \includegraphics[width=1\linewidth]{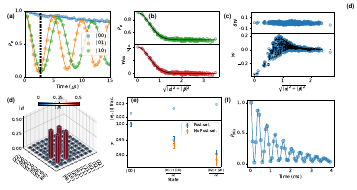}
    \caption{\textbf{NOON state tomography.} \textbf{(a)} Calibration of the wait time time for a generalized two-mode parity operation using transmon Ramsey experiments following the addition of a photon in each mode (here, modes three and five). The dashed black line indicates the chosen Ramsey time, which deviates from a parity measurement for both modes. The phases of the Ramsey fringes corresponding to a mode at that time is used to extract $\theta_1$ and $\theta_2$. \textbf{(b)} (Top) Calibration of the contrast scaling ($p_e,p_g$) required for error mitigation of the joint parity measurement. (Bottom) Wigner function of the two-mode vacuum state following rescaling, as a function of $\sqrt{\alpha^2+\beta^2}$. \textbf{(c)} Comparison of the measured data against the Wigner function at the corresponding points of the reconstructed density matrix. Their difference (top) is used to estimate an upper bound on the error in the Wigner measurement. \textbf{(d)} Reconstructed density matrix of a NOON state with $N=1$. \textbf{(e)} (Top) Fraction of shots not in $\ket{g}$ in the post-selection experiment, found to be $\leq 2\%$ up to $N=2$. (Bottom) Fidelity extracted from two-mode Wigner tomography for NOON states up to $N=2$, with and without post-selection. \textbf{(f)} Ramsey experiment for the coherence of $(\ket{10}+\ket{01})/\sqrt{2}$, resulting in a decay time of $1.27\pm \qty{0.05}{\milli\second}$.}
    \label{multimode wigner}
\end{figure*}
Multimode Wigner tomography is implemented using the pulse sequence shown in Figure 4(c) of the main text. In this procedure, both cavities are driven to a set of chosen displacements ($\alpha, \beta$), and we subsequently use a transmon Ramsey sequence to measure a generalized parity-like operator:
\begin{eqnarray}
    \widetilde{\Pi} &=& \cos\left(\theta_1 \hat{n}_1+\theta_2 \hat{n}_2\right) \\
    &=& \cos\left(\theta_1 \hat{n}_1\right)\cos\left(\theta_2 \hat{n}_2\right) - \sin\left(\theta_1 \hat{n}_1\right)\sin\left(\theta_2 \hat{n}_2\right), \nonumber
\end{eqnarray}
where $n_i$ is the photon number operator for mode $i$. The angles $\theta_1$ and $\theta_2$ are calibrated using Ramsey experiments following the addition of a photon into the respective modes, shown in Fig.~\ref{multimode wigner}(a). We calibrate the scale factor between the cavity drive amplitude and the cavity displacement for each mode with the Wigner function of the vacuum state, using the procedure detailed in Section~\ref{sec:wigner_tomography}. We calibrate the reduction in contrast from transmon decoherence during the parity measurement, due to the weak coupling, using the Wigner function of the two-mode vacuum $\ket{00}$. We fit the transmon population as a function of the distance of the two-mode Wigner function of the vacuum state from the origin ($\sqrt{\alpha^2+\beta^2}$), to extract $p_e$ and $p_g$, similar to Eqn.~\ref{Wigner0_calibration}, as shown in Fig.~\ref{multimode wigner}(b). This procedure allows us to extract the Wigner functions for two-mode states at a set of chosen cavity displacements for each mode ($\mathcal{W}(\alpha_i,\beta_j)$), which we invert to extract the two-mode density matrix using the procedure detailed in~\cite{chakram2020multimode}. The absolute value of the density matrix ($|\rho|$) for the entangled Bell state $\frac{\ket{10}+\ket{01}}{\sqrt{2}}$ is shown in Fig.~\ref{multimode wigner} (c). 

We perform two-mode Wigner tomography both with and without post-selection for the vacuum and NOON states up to $N=2$; the results are shown in Fig.~\ref{multimode wigner}(e). We achieve fidelities of $0.95\pm0.011$ ($0.934\pm0.012$) and $0.905\pm0.012$ ($0.887\pm0.016$) with (without) post-selection for NOON states with $N=1$ and $N=2$, respectively. The improvement in fidelity with post-selection is consistent with the excited-state fraction shown at the top of Fig.~\ref{multimode wigner}(e). We estimate the error in our reconstructed fidelity by extending the methods presented in Section~\ref{wigner error} to the two-mode case and assessing how measurement and calibration noise translate to fidelity error. Specifically, we estimate measurement noise by comparing the measured Wigner function with that of the reconstructed $\rho$ at corresponding phase-space points; the results for the $N=1$ NOON state are shown in Fig.~\ref{multimode wigner}(c). The deviation $\langle\delta\mathcal{W}\rangle$ is comparable to that observed in the single-mode case. We inject Gaussian random noise with the same standard deviation and perform the reconstruction multiple times (10 iterations) to estimate the fidelity error.

\subsection{NOON state coherence using encode and decode}
In place of multimode Wigner tomography, we employ a faster method to characterize the state within the NOON subspace—after verifying via photon-number-resolved spectroscopy that less than 2\% of the population lies outside this space. The state is characterized by performing qubit tomography following decoding of the NOON state. By idling in the NOON state between the encoding and decoding operations, we can measure its coherence time by sandwiching the encode, idle, and decode operations within a transmon \(ge\) Ramsey experiment. The result of this experiment (after resetting to remove measurement errors from populations outside the subspace) is shown in Fig.~\ref{multimode wigner}(f), corresponding to a coherence time of $1.27\pm \qty{0.05}{\milli\second}$ for the $\frac{\ket{10}+\ket{01}}{\sqrt{2}}$ Bell state.

\section{Data normalization}\label{Data normalization}

Probability normalization for transmon measurements was performed using a reference set of single-shot IQ data, where the transmon was prepared in $\ket{g}$ and $\ket{e}$ after each experiment. A linear rotation transformation was applied to the IQ data to maximize the separation between the rotated median IQ values for $\ket{g}$ and $\ket{e}$ ($\text{I}^\text{med}\text{g}$ and $\text{I}^\text{med}\text{e}$, respectively) along the I quadrature. Subsequent transmon measurement IQ data were rotated using the same transformation and then normalized along the I quadrature to extract the probabilities using $P_e = (\text{I}_\text{meas}-\text{I}^\text{med}_\text{g}) / (\text{I}^\text{med}_\text{e} - \text{I}^\text{med}_\text{g})$. 

Photon-number-resolved qubit spectroscopy (PNRQS) datasets were rescaled to account for SPAM errors arising from decay of the transmon $\ket{e}$ state during the long resolved qubit pulse, which is around $\qty{8}{\micro\second}$ long. The scaling was calibrated using PNRQS of the $\ket{0}$ state of the cavity, scaling the measured transmon population to 1. 

Datasets for the $\pi_{ef}$ and $\ket{f0}$–$\ket{g1}$ sideband pulse trains, the $\ket{f0}$–$\ket{g1}$ Rabi data in Fig. \ref{fig:pnr_sideband} and Fig. \ref{fig:rabi_fidelities}, and the NOON state Ramsey experiment in Fig. \ref{multimode wigner}(f) were rescaled to ensure that the initial population at zero pulses or time was 1. This correction mitigates SPAM errors introduced during the reset of transmon states outside the interaction subspace being measured, which was performed before the final transmon measurement.

The reset was implemented using an $\ket{f0_r}$–$\ket{g1_r}$ sideband to the readout mode for $\qty{6}{\micro\second}$, leading to decay within the interaction subspace under measurement. Corrections for SPAM errors in Wigner tomography due to this effect are detailed in Section \ref{sec:wigner_tomography}.

\bibliography{references}